\documentclass[12pt]{article}
\usepackage{amsmath,amssymb,exscale,psfrag,epsfig,axodraw,color}
\usepackage{hyperref}
\input epsf
\newcommand{\Dslash}{{\not \!\!D}}

\newcommand{\Xslash}{{\not \!\!X}}
\newcommand{\partialslash}{{\not \!\partial}}
\newcommand{\dslash}{{\not \!d}}
\newcommand{\eslash}{{\not \!e}}
\newcommand{\ddh}{{\partial h}}
\newcommand{\bR}{{\bf R}}
\newcommand{\br}{{\bf r}}
\newcommand{\bs}{{\bf s}}

\begin{document}
\topmargin -1.0cm
\oddsidemargin -0.8cm
\evensidemargin -0.8cm

\thispagestyle{empty}

\vspace{40pt}

\begin{center}
\vspace{40pt}

\Large \textbf{Anarchy with linear and bilinear interactions}

\end{center}

\vspace{15pt}
\begin{center}
{\large Leandro Da Rold} 

\vspace{20pt}

\textit{Centro At\'omico Bariloche, Instituto Balseiro and CONICET}
\\[0.2cm]
\textit{Av.\ Bustillo 9500, 8400, S.\ C.\ de Bariloche, Argentina}

\end{center}

\vspace{20pt}
\begin{center}
\textbf{Abstract}
\end{center}
\vspace{5pt} {\small \noindent
Composite Higgs models with anarchic partial compositeness require a scale of new physics ${\cal O}(10-100)$~TeV, with the bounds being dominated by the dipole moments and $\epsilon_K$. The presence of anarchic bilinear interactions can change this picture. We show a solution to the SM flavor puzzle where the electron and the Right-handed quarks of the first generation have negligible linear interactions, and the bilinear interactions account for most of their masses, whereas the other chiral fermions follow a similar pattern to anarchic partial compositeness. We compute the bounds from flavor and CP violation and show that  neutron and electron dipole moments, as well as $\epsilon_K$ and $\mu\to e\gamma$, are compatible with a new physics scale below the TeV. $\Delta F=2$ operators involving Left-handed quarks and $\Delta F=1$ operators with $d_L$ give the most stringent bounds in this scenario. Their Wilson coefficients have the same origin as in anarchic partial compositeness, requiring the masses of the new states to be larger than ${\cal O}(6-7)$~TeV.}

\vfill
\noindent
\eject


\section{Introduction}
Composite Higgs models where the Higgs arises as a resonance of a strongly coupled field theory (SCFT) at the TeV scale offer a very interesting solution to the hierarchy problem. The realization of flavor in these models is very different from the Standard Model (SM). One of the most attractive ideas in this context has been anarchic partial compositeness, where the SM fermions have linear interactions with operators of the SCFT, generating at low energies mixing with the fermionic resonances and masses for the SM fermions~\cite{Kaplan:1991dc}. Although the SCFT is flavor anarchic, the mixings can be hierarchical if the scale $\Lambda_{UV,\psi}$ at which the interactions are generated is much larger than the TeV and the operators interacting linearly with the SM fermions have different anomalous dimensions. A very attractive feature of partial compositeness is that the hierarchical mixing can simultaneously reproduce the hierarchical spectrum of SM fermions as well as the angles of the CKM matrix. Moreover, flavor violating processes are suppressed by the same small mixings, leading to non-generic Wilson coefficients of flavor violating higher-dimensional operators~\cite{Grossman:1999ra,Huber:2000ie,Gherghetta:2000qt,Contino:2004vy,Agashe:2004cp}. Although this mechanism is enough to satisfy most of the bounds, $\epsilon_K$ requires the scale of resonances $m_\rho\gtrsim{\cal O}(10)$~TeV, whereas $\mu\to e\gamma$, as well as the electron and neutron dipole moments require $m_\rho/g_\rho\gtrsim{\cal O}(5-40)$~TeV, where $g_\rho$ is the coupling between resonances~\cite{Agashe:2006iy,Csaki:2008zd,Bauer:2009cf}. These bounds generate a problem of fine-tuning of the electroweak (EW) scale.

There are several interesting alternatives in the literature. One example corresponds to the introduction of flavor symmetries in the SCFT~\cite{Fitzpatrick:2007sa,Cacciapaglia:2007fw,Csaki:2008eh,Santiago:2008vq,Redi:2011zi,Barbieri:2012uh}, as well as full compositeness of the Left- or Right-handed SM quarks~\cite{Redi:2011zi}. In these models the only non-trivial flavor structure is proportional to the SM Yukawa coupling, thus bounds from flavor violation can be satisfied with $m_\rho\sim$~TeV. Besides the fact that in this case the idea of anarchy is abandoned, a new problem emerges: if the light quarks are fully composite, 4-fermion operators are only suppressed by the TeV scale, inducing deviations in dijet production at LHC that have not been observed~\cite{Domenech:2012ai}. The authors of Ref.~\cite{Bauer:2011ah} have proposed a symmetry to supress the main contributions to $\epsilon_K$ in anarchic partial compositeness, however Ref.~\cite{DaRold:2012sz} has shown that the suppression is spoiled once quark masses are taken into account. Ref.~\cite{DaRold:2017dbr} has provided a completion of the set-up of~\cite{Bauer:2011ah}, discussing to which extent the bounds from $\epsilon_K$ can be alleviated, and showing that dipole operators can be suppressed.

In most of the analysis of models with partial compositeness, bilinear interactions of the SM fermions with the SCFT have been neglected. One of the main reasons for this being that their effect is considered very small, due to the large anomalous dimension of the Higgs operator ${\cal O}_H$~\footnote{In order to solve the hierarchy problem, the lowest dimensional operator of the SCFT that is a gauge singlet and can create two Higgses at low energies, 
must have anomalous dimension $\gtrsim 4$. As shown in Ref.~\cite{Rattazzi:2008pe}, under some general assumptions, this implies that ${\rm dim}\ {\cal O}_H$ has to be significantly larger than 1.} and the huge separation between the TeV scale and the scale $\Lambda_{UV,H}$ at which the bilinear interactions are generated. In this paper we want to pursue the idea of flavor anarchy by adding, to the usual scenario of anarchic partial compositeness, the effect of bilinear interactions with couplings that are anarchic at the scale $\Lambda_{UV,H}$. As is well known, anarchic bilinear interactions by their own can not solve the flavor puzzle of the SM. However, they can provide masses to the light fermions, as those of the first generation, and, when combined with linear interactions, they offer new possibilities for flavor~\cite{Agashe:2008fe}. In particular, since the most stringent bounds in anarchic partial compositeness arise from processes involving the fermions of the first generation,
taking some chiralities of these fermions mostly elementary those bounds can be relaxed. We will show a solution where, in an anarchic SCFT, the interplay of linear and bilinear interactions can lead to a realistic model, and the most stringent bounds can be alleviated. 

Some recent works have also considered the presence of bilinear interactions, as Ref.~\cite{Matsedonskyi:2014iha} that has briefly discussed them in scenarios with flavor symmetries, and Refs.~\cite{Cacciapaglia:2015dsa,Panico:2016ull}. However the UV scales and the spectrum of resonances of these models, as well as the parametric dependence of the bounds, are different from the approach that we propose here. Ref.~\cite{Agashe:2008fe} considered the presence of linear and bilinear interactions for the lepton sector, although the authors did not discuss the bounds from flavor observables. 

The paper is organized as follows: in sec.~\ref{sec-scenario} we describe a model containing linear and bilinear interactions of the SM fermions with operators of a flavor anarchic SCFT. We show a solution that leads to the hierarchies of the SM fermion masses and to the pattern of mixing in the quark and lepton sectors. In sec.~\ref{sec-flavor-cp-v} we compute the contributions to the Wilson coefficients of the operators that provide the main sources of flavor and CP violation. We compare the bounds with the case of anarchic partial compositeness and show how some of the most stringent bounds can be alleviated. Our conclusions are in sec.~\ref{sec-conclusions}. In the Appendices~\ref{ap-minimal-model} and \ref{ap-MCHM5} we consider effective models of a composite Higgs. In Ap.~\ref{ap-minimal-model} we present a minimal model, described in terms of a two-site theory, with a composite Higgs and set of fermionic and spin 1 resonances. In Ap.~\ref{ap-MCHM5} we consider a pseudo Nambu-Goldstone boson (pNGB) Higgs arising from SO(5)/SO(4), with a set of fermionic resonances. In both cases we compute the Wilson coefficients of the operators of sec.~\ref{sec-flavor-cp-v}.

\section{Flavor with linear and bilinear couplings}\label{sec-scenario}
We consider a theory that contains a flavorful strongly interacting sector that confines and generates resonances with masses $m_\rho\sim$~few TeV. As we specify below, the flavor structure of the SCFT is assumed to be anarchic. The strong dynamics also produces a light spinless state, that can be identified with the Higgs boson.~\footnote{As has been extensively discussed in the literature, a very interesting possibility to suppress the Higgs mass compared with $m_\rho$ is to have a Higgs as a pNGB of the SCFT~\cite{Contino:2003ve,Agashe:2004rs}. In Ap.~\ref{ap-MCHM5} we show a low energy effective model containing a pNGB Higgs.} There is also an elementary sector, external to the SCFT, containing the SM fermions. The SM gauge bosons are also elementary, and weakly gauge a subgroup of the global symmetry of the SCFT.

We assume that at the high scale $\Lambda_{UV,H}\gg m_\rho$ the SM quarks couple bilinearly to an operator ${\cal O}_H$, that is the SCFT operator with lowest dimension that at low energies can create a Higgs: 
\begin{equation}\label{LUV1}
{\cal L}\supset
\sum_{\psi=u,d}\frac{\omega_{y^\psi}^{jk}(\Lambda_{UV,H})}{\Lambda_{UV,H}^{\Delta_H-1}}\bar q_L^j\psi_R^k{\cal O}_H + {\rm h.c.}\ ,
\end{equation}
where $\Delta_H$ is the scaling dimension of ${\cal O}_H$ in energy units, and $j,k$ label generations. In Technicolor ${\cal O}_H$ is an operator bilinear in techniquarks, with $\Delta_H\simeq 3$. In the present scenario we do not need to specify the content of ${\cal O}_H$, besides $\Delta_H$ and $\Lambda_{UV,H}$ are taken as free parameters. Anarchy dictates that the couplings at the UV scale: $\omega_\psi^{jk}(\Lambda_{UV,\psi})$, have no structure at all, thus all the coefficients of these matrices are of the same order. Besides the interactions of Eq.~(\ref{LUV1}), at the scale $\Lambda_{UV,H}$ higher dimensional operators providing sources of flavor and CP violation can also be generated. As we will discuss in sec.~\ref{sec-flavor-cp-v}, to avoid unacceptably large corrections, $\Lambda_{UV,H}$ has to be taken larger than ${\cal O}(10^5)$~TeV~\cite{Rattazzi:2008pe}. 

We also assume that at a scale $\Lambda_{UV,\psi}\gg m_\rho$ the SM quarks couple linearly to operators of the SCFT ${\cal O}_\psi$
\begin{equation}\label{LUV2}
{\cal L}\supset\sum_{\psi=q_L,u_R,d_R}\frac{\omega_\psi^{jk}(\Lambda_{UV,\psi})}{\Lambda_{UV,\psi}^{\Delta_\psi^k-5/2}}\bar\psi^j{\cal O}_\psi^k
\ ,
\end{equation}
where $\Delta_\psi$ is the scaling dimension of ${\cal O}_\psi$. We assume that $\Lambda_{UV,\psi}$ is the same for all the SM quarks, it does not need to be equal to $\Lambda_{UV,H}$, although {\it a priori} there is no need to take them different. $\Delta_\psi$ can take different values for the different flavors and generations. As for the bilinear couplings, in the anarchic scenario there is no flavor structure for the linear couplings at the scale $\Lambda_{UV,\psi}$, therefore all the coefficients of $\omega_\psi(\Lambda_{UV,\psi})$ are of the same order. For simplicity we assume that each SM fermion have linear interactions with just one operator of the SCFT.

At low energies the composite operators ${\cal O}_\psi$ create fermionic resonances. We assume that the masses of the lowest lying level of resonances are of the same order, $m_\rho$, for all the operators ${\cal O}_\psi^k$, and that the couplings among the resonances can be described in terms of a single parameter $1<g_\rho\lesssim 4\pi$. We find it useful to define the scale $f= m_\rho/g_\rho$. 
Since EW precision tests require $v/f\lesssim 0.3-0.5$~\cite{Grojean:2013qca}, we will assume this bound on $f$ to be satisfied in the rest of the paper.

We assume that the energy dependence of the couplings $\omega_{y^\psi}$ and $\omega_\psi$ is driven by the dimension of the corresponding operators. Up to corrections ${\cal O}(1)$, at the scale $m_\rho$ the linear and bilinear couplings can be approximated by:
\begin{equation}\label{eq-lowEw}
y_\psi^{jk}\sim \omega_{y^\psi}^{jk}(\Lambda_{UV,H})\left(\frac{m_\rho}{\Lambda_{UV,H}}\right)^{\Delta_H-1} \ ,
\qquad
\lambda_\psi^{jk}\sim \omega_\psi^{jk}(\Lambda_{UV,\psi})\left(\frac{m_\rho}{\Lambda_{UV,\psi}}\right)^{\Delta_\psi^k-5/2} \ .
\end{equation}
As usual in composite Higgs models, for large separation of scales and $\Delta_H$ significantly larger than 1, $y$ is strongly suppressed, in such a way that the Yukawa coupling can not account for the mass of the top, however it can give a significant contribution to the mass of the light quarks. The coupling $\lambda_\psi$ is responsible for partial compositeness of the SM quarks. Taking $\Delta_\psi$ near 5/2 for $q_L^3$ and $u_R^3$, the top becomes significantly composite, and an ${\cal O}(1)$ Yukawa can be generated through this mechanism even for a large separation of scales. On the other hand, for the Right-handed bottom, as well as for the quarks of the first and second generation, $\Delta_\psi$ will be taken flavor dependent and larger than 5/2, leading to hierarchical mixings.

To proceed with the analysis, we find it convenient to rotate to the basis where $\lambda_\psi$ are diagonal~\cite{Panico:2015jxa}:
\begin{equation}
\lambda_\psi=U_\psi^\dagger \lambda_\psi^D U_{{\cal O}_\psi} \ ,
\end{equation}
where $U_\psi$ rotates the elementary fermions and $U_{{\cal O}_\psi}$ the composite operators. If the dimensions $\Delta_\psi^k$ differ for different generations, the eigenvalues of $\lambda_\psi$ are hierarchical: $\lambda_\psi^j/\lambda_\psi^k\sim(m_\rho/\Lambda_{UV})^{\Delta_\psi^j-\Delta_\psi^k}$, and the coefficients of $U_{{\cal O}_\psi}$ are of order: $U_{{\cal O}_\psi}^{jk}\sim\lambda_\psi^j/\lambda_\psi^k$, for $\Delta_\psi^j>\Delta_\psi^k$. On the other hand, for anarchic $\omega$ all the coefficients of the matrix $U_\psi$ are of the same order. 
Since to leading order the evolution of the bilinear coupling is driven by $\Delta_H$, in the new basis the matrix of Yukawa couplings $y_\psi$ is also anarchic. In the following we will work in this basis.

At energies below $m_\rho$ the fermionic states can be integrated-out, generating the following effective Yukawa couplings for the elementary fermions
\begin{equation}\label{Yjk}
Y_\psi^{jk}\simeq y_\psi^{jk} + c_\psi^{jk}\frac{\lambda_q^j\lambda_\psi^k}{g_\rho}
\end{equation}
with $c_\psi^{jk}\sim{\cal O}(1)$ anarchic coefficients, arising from the Yukawa couplings of the composite states. The first term is the usual coupling of theories of Technicolor and the second one is the Yukawa generated in partial compositeness.

In the next sections we will sometimes use $\lambda_L$ and $\lambda_R$ for the mixing of the Left- and Right-handed fermions, such that for quarks $\lambda_L=\lambda_q$ and $\lambda_R=\lambda_u,\lambda_d$, whereas for the leptons $\lambda_L=\lambda_\ell$ and $\lambda_R=\lambda_e$. Generation indices will be added whenever necessary.

\subsection{Quark masses and mixing angles}
The masses and mixing of the SM quarks are obtained by bi-unitary diagonalization of $Y$, defined in Eq.~(\ref{Yjk}). We will consider the situation in which the bilinear coupling at the scale $m_\rho$ is of the same order as the Yukawa couplings of the SM fermions of the first generation:
\begin{equation}\label{yjk}
y_\psi^{jk}\sim y_\psi^{SM} \ ,\qquad \psi=u,d \ , 
\end{equation}
thus $y_\psi^{jk}\sim {\rm MeV}/v$. Notice that once the value of the Yukawa coupling at certain scale is fixed, Eq.~(\ref{eq-lowEw}) gives a relation between $\Delta_H$ and $\Lambda_{UV,H}$. 

We will consider the case of large dimensions $\Delta_u^1$ and $\Delta_d^1$, with $\lambda^1_{u,d}\ll y_{u,d}g_\rho/\lambda^j_q$, and in our calculations we will take:
\begin{equation}\label{lambdaud}
\lambda^1_{u,d}=0 \ , 
\end{equation}
neglecting, as a leading order approximation, the linear mixing of the Right-handed quarks of the first generation. 

Since $y_\psi$ is too small to generate the masses of the fermions of the second and third generation, those masses must be generated by partial compositeness. Using Eqs.~(\ref{yjk}) and (\ref{lambdaud}) in (\ref{Yjk}), the resulting Yukawa matrix can be approximated by:
\begin{equation}
Y_\psi\sim\left(\begin{array}{ccc}y_\psi&\lambda_q^1\lambda_\psi^2/g_\rho&\lambda_q^1\lambda_\psi^3/g_\rho\\y_\psi&\lambda_q^2\lambda_\psi^2/g_\rho&\lambda_q^2\lambda_\psi^3/g_\rho\\y_\psi&\lambda_q^3\lambda_\psi^2/g_\rho&\lambda_q^3\lambda_\psi^3/g_\rho\end{array}\right) \ ,
\end{equation}
where we have not shown factors of ${\cal O}(1)$, and where we have neglected $y^{jk}$ compared with $\lambda_q^j\lambda_\psi^k/g_\rho$ for $k=2,3$.

The CKM matrix of EW charged interactions can be obtained by taking $\lambda_q^j$ as in the case of anarchic partial compositeness: 
\begin{equation}\label{lambdaCKM}
\lambda^1_q/\lambda^2_q\sim\lambda_C\ , \qquad \lambda^2_q/\lambda^3_q\sim\lambda_C^2 \ ,
\end{equation}
where $\lambda_C\simeq 0.22$ is the Cabibbo angle. The SM quark masses can be reproduce if Eq.~(\ref{yjk}) and the following relations are satisfied:
\begin{align}
&\lambda_u^3\sim y_t^{\rm SM} \frac{g_\rho}{\lambda_q^3} \ ,
\qquad
\lambda_u^2\sim y_c^{\rm SM} \frac{g_\rho}{\lambda_q^2} \ , \nonumber \\
&\lambda_d^3\sim y_b^{\rm SM} \frac{g_\rho}{\lambda_q^3} \ ,
\qquad
\lambda_d^2\sim y_s^{\rm SM} \frac{g_\rho}{\lambda_q^2} \ .
\label{mixqR}
\end{align}
As usual in partial compositeness, the top mass requires a large degree of compositeness of both chiralities.

Using Eqs.~(\ref{yjk}), (\ref{lambdaud}), (\ref{lambdaCKM}) and (\ref{mixqR}) in (\ref{Yjk}), it is straightforward to check that there is one eigenvalue of order $y_\psi$, reproducing the masses of the fermions of the first generation, and two eigenvalues of order $\frac{\lambda_q^j\lambda_\psi^j}{g_\rho}\sim \frac{m_\psi^j}{v}$ for $j=2,3$ that give the masses of the quarks of the second and third generation. It is also straightforward to show that, to leading order in powers of $y_\psi$ and $\lambda_R$, $Y_\psi Y_\psi^\dagger$ has exactly the same structure as in the case of anarchic partial compositeness.

\subsection{Leptons}\label{sec-leptons}
For the charged leptons we will consider a scenario similar to the quarks, with linear and bilinear interactions for doublets $\ell_L^j$ and singlets $e_R^j$. The size of $y_e$ is controlled by the running of $\omega_{y^e}$, Eq.~(\ref{eq-lowEw}), as in the sector of quarks. The scale $\Lambda_{UV,\psi}$ for leptons will be taken equal to the scale for quarks. The Yukawa coupling of the charged leptons can be obtained from Eq.~(\ref{Yjk}), taking $\psi=e$ and changing $q$ by $\ell$. Following Eq.~(\ref{yjk}), the bilinear interactions give a contribution to the mass of the charged leptons of order MeV, and thus are able to account for the mass of the electron, whereas the masses of the muon and tau require partial compositeness. 
We will consider the case of large dimensions $\Delta_\ell^1$ and $\Delta_e^1$, with$(\lambda^1_{\ell,e})^2\ll y_e g_\rho$, and in practice we will take:
\begin{equation}\label{lambdaelle}
\lambda^1_\ell=\lambda^1_e=0 \ , 
\end{equation}
neglecting the linear mixing of the Left and Right-handed electron. 

For the second and third generations we choose hierarchical couplings of same size for both chiralities~\cite{Redi:2013pga}:
\begin{equation}
\lambda_\ell^j\sim\lambda_e^j\sim\sqrt{\frac{m_{e^j} g_\rho}{v}}\ , \qquad j=2,3 \ .
\end{equation}
With this choice of parameters the unitary matrices that diagonalize $Y_e$ have small mixing angles, and the PMNS matrix is approximately determined by the unitary matrix diagonalizing the neutrino sector.

For the neutrinos we will follow the proposal of Refs.~\cite{Agashe:2008fe} and~\cite{Panico:2015jxa,Panico:2016ull}, where neutrino masses are generated by a bilinear operator involving the Left-handed leptons only, the neutrinos being Majorana fermions. We add to Eqs.~(\ref{LUV1}) and~(\ref{LUV2}) the dimension five Weinberg operator:
\begin{equation}
{\cal L}=
\frac{\omega_{y^\nu}^{jk}(\Lambda_{UV,H})}{\Lambda_{UV,H}^{\Delta'_{H^2}-1}} (\bar\ell_L^j)^c\ell_L^k{\cal O}'_{H^2} \ ,
\label{LUVnu}
\end{equation}
where ${\cal O}'_{H^2}$ is the lowest dimensional operator of the SCFT that is a triplet of SU(2)$_L$ with hypercharge $Y=1$, and $\omega_{y^\nu}(\Lambda_{UV,H})$ is an anarchic coupling. At low energies, Eq.~(\ref{LUVnu}) generates masses and mixings for the neutrinos. Up to corrections ${\cal O}(1)$, the coupling at the energy scale $m_\rho$ is:
\begin{equation}\label{eq-lowEwnu}
y_\nu^{jk}\simeq \omega_{y^\nu}^{jk}(\Lambda_{UV,H})\left(\frac{m_\rho}{\Lambda_{UV,H}}\right)^{\Delta'_{H^2}-2} \ ,
\end{equation}
such that for $\Delta'_{H^2}>2$ and $\Lambda_{UV,H}\gg m_\rho$ small neutrinos masses can be naturally obtained: $m_\nu\sim y_\nu v^2/\Lambda_{UV,H}$. If the scale $m_\rho$ and the size of $\omega_{y^\nu}$ are fixed, reproducing the neutrino masses requires a relation between $\Lambda_{UV,H}$ and $\Delta'_{H^2}$. If, as discussed in sec.~\ref{sec-flavor-cp-v}, $\Lambda_{UV,H}\gtrsim 10^5$ TeV, taking $\omega_{y^\nu}\sim{\cal O}(\pi)$ and $m_\rho\sim{\cal O}(3\ {\rm TeV})$, we obtain the bound $\Delta'_{H^2}\lesssim 3.6$.

For anarchic Yukawa all the neutrino masses are naturally of the same order and the mixing angles are ${\cal O}(1)$. Therefore the large mixing angles of the PMNS matrix can be generated by Eq.~(\ref{LUVnu}).

\section{Flavor and CP violation}\label{sec-flavor-cp-v}
We study in this section the main sources of flavor and CP violation when linear and bilinear couplings are present. The Wilson coefficients of the corresponding operators have a contribution arising from partial compositeness, that have been extensively studied in the literature~\cite{Csaki:2008zd,Bauer:2009cf,KerenZur:2012fr}, and there are new contributions induced by the bilinear interactions. For the choice of parameters of Eqs.~(\ref{yjk}), (\ref{lambdaud}) and (\ref{lambdaelle}), processes involving Right-handed quarks of the first generation, as well as electrons, require insertions of the bilinear interactions, therefore the corresponding Wilson coefficients are dominated by contributions that have a different structure compared with partial compositeness. We study the bounds arising from these processes and compare them with the predictions of anarchic partial compositeness. We also briefly comment which are the most stringent bounds in this scenario.

We stress that we do not attempt to make a complete analysis of flavor in composite Higgs models. Instead we consider the largest flavor violating effects.

We will consider the flavor and CP violating operators contained in the following Lagrangian:
\begin{equation}\label{Lflavor}
{\cal L}= C_{d'}^{ab} {\cal Q}_{d'}^{ab}+C_p^{ab}{\cal Q}_p^{ab}+C_{4f}^{abcd}{\cal Q}_{4f}^{abcd}\ ,
\end{equation}
where $a,b,\dots$ collect flavor and generation indices, we will sometimes use these indices to denote generations: $a=1,2,3$, and when necessary we will use them to denote flavor and chirality: $a=d_L,s_L,\dots$ The operators in Eq.~(\ref{Lflavor}) are given by:~\footnote{The SM field H can be identified with the operator ${\cal O}_H$ at the low energy scale $m_\rho$ after proper normalization.}
\begin{align}
&{\cal Q}_{d'}^{ab}=g_{\rm SM}\bar \psi^a F_{\mu\nu} H \sigma^{\mu\nu}\psi^b  \ , \label{operatorQdp} \\
&{\cal Q}_p^{ab}=\bar \psi^a \gamma^{\mu} \psi^b H^\dagger \overleftrightarrow D_\mu H  \ , \label{operatorQp} \\
&{\cal Q}_{4f}^{abcd}=\bar \psi^a\gamma^\mu\psi^b \bar \psi^c \gamma_\mu \psi^d \ . \label{operatorQ4f}
\end{align}
After EWSB we find it useful to define the operators:
\begin{align}
&{\cal Q}_d^{ab}=g_{\rm SM} m_b \bar \psi^a F_{\mu\nu}\sigma^{\mu\nu}\psi^b  \ , \label{operatorQd} \\
&{\cal Q}_Z^{ab}=Z_\mu\bar \psi^a \gamma^{\mu} \psi^b  \ , \label{operatorQZ} 
\end{align}
where the corrections to the $Z$-interactions arise from penguin operators by putting the Higgs to its vacuum expectation value.

In the presence of linear and bilinear interactions, the Wilson coefficients of the operators~(\ref{operatorQ4f}), (\ref{operatorQd}) and (\ref{operatorQZ}), expanding to second order in $\lambda$ and $y$, can be estimated as:
\begin{align}
&C_d^{ab} \sim \frac{1}{m_b}\frac{1}{(4\pi)^2}\frac{v}{m_\rho^2g_\rho}\left[g_\rho^2\lambda_L\lambda_R +g_\rho y(\lambda_L^2+\lambda_R^2)+ y^2\lambda_L\lambda_R\right]^{ab}\ , \label{Cd} \\
&C_{Z_{L,R}}^{ab} \sim \frac{g}{c_W}\frac{v^2}{m_\rho^2g_\rho^2}(g_\rho^2\lambda_{L,R}^2+g_\rho y\lambda_L\lambda_R+y^2\lambda_{R,L}^2)^{ab}  \ , \label{CZ} \\
&C_{4f}^{abcd} \sim \frac{1}{m_\rho^2g_\rho^2} [\lambda^4(1+g_\rho y\frac{v^2}{m_\rho^2}+y^2\frac{v^2}{m_\rho^2})]^{abcd} \ , \label{C4f} 
\end{align}
where we have assumed that the dipole operators are effects of order 1-loop and we have specified the chirality of the $Z$ coupling. We have not specified the indices $L$ or $R$ of the factors $\lambda$ in Eq.~(\ref{C4f}) because they depend on the values of the superindices of $C_{4f}^{abcd}$. In the Appendices~\ref{ap-minimal-model} and \ref{ap-MCHM5} we compute the predictions for these Wilson coefficients in two effective models of composite Higgs, and we show that they are in agreement with the estimates of Eqs. (\ref{Cd})-(\ref{C4f}).

The flavor and generation indices of the Wilson coefficients: $a,b,\dots$, are inherited from $\lambda_\psi^a$ and $y_\psi^{ab}$.
In Eqs.~(\ref{Cd}) and (\ref{CZ}), to zeroth order in $y$ there are only insertions of $\lambda$, thus $(\lambda^2)^{ab}=\lambda^a\lambda^b$, whereas in the structure of ${\cal O}(y)$ one of the indices arises from $\lambda$ and the other from $y$, for example: $(y\lambda^2)^{ab}=\lambda^ay^{bc}\lambda^c+\dots$, where $c$ runs over generations and the dots stand for permutations of indices satisfying the aforementioned conditions. Since $y$ is anarchic, the size of $y^{bc}$ is independent of the value of the indices, on the other hand, being $\lambda$ hierarchical, $\lambda^c$ is dominated by the third generation, $c=3$. For the terms quadratic in $y$, $(y^2\lambda^2)^{ab}\sim\lambda^a y^{bc}y^{cd}\lambda^d+y^{ac}y^{bd}\lambda^c\lambda^d+\dots$, with $c,d$ running over generations, such that the flavor indices of the Wilson coefficients can be associated to two Yukawa couplings or to one index of the Yukawa and one of $\lambda$. Similar considerations apply to Eq.~(\ref{C4f}), to zeroth order in $y$ each index of $C_{4f}$ is associated to the index of one $\lambda$. To ${\cal O}(y)$, one index of $C_{4f}$ is associated to one index of $y$, and the other three indices to three factors of $\lambda$, leaving free a fourth factor of $\lambda$ that is dominated by the third generation, with a chirality that accounts for the chiral flip induced by the insertion of $y$. To ${\cal O}(y^2)$, the indices of $C_{4f}$ can be associated to two Yukawa couplings and two factors of $\lambda$, or to one index of the Yukawa and three factors of $\lambda$, leaving free two or one factor of $\lambda$, respectively. The chirality of these factors account for the chiral flips induced by the insertions of $y$.

In partial compositeness $y=0$ and only the first term of Eqs.~(\ref{Cd})-(\ref{C4f}) is present. The effect of the bilinear interactions becomes relevant if $\lambda_L$ and/or $\lambda_R$ are very small, in which case the terms of order $y$ or $y^2$ can dominate. In our approach, these effects are relevant only for operators with $u_R$, $d_R$, $e_R$ and $e_L$, since for these fermions we consider $\lambda=0$. Thus the Wilson coefficients of the operators involving these fermions differ from the case of anarchic partial compositeness. We find that, for quarks, the terms of ${\cal O}(y)$ dominate over the terms of ${\cal O}(y^2)$. For the leptons, in operators involving $e_L$ and $e_R$, as electromagnetic dipole moment (EDM), the terms of ${\cal O}(y)$ are very small and those of ${\cal O}(y^2)$ dominate.

The operators of Eqs.~(\ref{operatorQdp})-(\ref{operatorQ4f}) can also be generated at the scale $\Lambda_{UV,H}$, with Wilson coefficients: 
\begin{equation}\label{CLambdaUVH}
C_{d'}\sim \frac{\omega_{d'}(\Lambda_{UV,H})}{\Lambda_{UV,H}^{\Delta_H+1}} \ ,
\qquad
C_p\sim \frac{\omega_p(\Lambda_{UV,H})}{\Lambda_{UV,H}^{\Delta_{H^2}}} \ ,
\qquad
C_{4f}\sim \frac{\omega_{4f}(\Lambda_{UV,H})}{\Lambda_{UV,H}^2} \ .
\end{equation}
We take the dimensionless couplings $\omega$, at the scale $\Lambda_{UV,H}$, anarchic and of size $1\lesssim\omega(\Lambda_{UV,H})\lesssim 4\pi$. At the low energy scale $m_\rho$, these couplings can be approximated by:
\begin{equation}
\omega_{d'}(m_\rho)\sim \omega_{d'}(\Lambda_{UV,H}) \left(\frac{m_\rho}{\Lambda_{UV,H}}\right)^{\Delta_H-1} \ ,
\qquad
\omega_p(m_\rho)\sim \omega_p(\Lambda_{UV,H}) \left(\frac{m_\rho}{\Lambda_{UV,H}}\right)^{\Delta_{H^2}-2} \ .
\end{equation}

In order to estimate the size of these contributions we assume $\omega(\Lambda_{UV,H})\sim{\cal O}(\pi)$. Demanding $C_{4f}\gtrsim (2\times 10^5\ {\rm TeV})^{-2}$, as required by the $K^0-\bar K^0$ system, we obtain $\Lambda_{UV,H}\gtrsim 3\times 10^5\ {\rm TeV}$. Using this bound and Eq.~(\ref{yjk}), we obtain: $\Delta_H\lesssim 2.1$. Taking these bounds for $\Lambda_{UV,H}$ and $\Delta_H$, the contributions to flavor and CP violating observables from Eq.~(\ref{CLambdaUVH}) can be safely neglected. We will assume this to be the case in the rest of the paper.

Dangerous $\Delta F=2$ operators can also be generated by tree-level Higgs exchange. These corrections are suppressed if the Higgs is a pNGB and the fermions transform under suitable representations of the global symmetry of the strong sector~\cite{Agashe:2009di,Mrazek:2011iu}, for that reason we will not consider them in our analysis. In Ap.~\ref{ap-MCHM5} we show an effective model where the conditions for the alignment are realized.

\subsection{Quarks}
We consider the effects of flavor and CP violation in the quark sector. We define:
\begin{equation}
\delta_q=\left(\frac{\lambda_q^3}{g_\rho}\right)^2 \lambda_C^3 \simeq 10^{-2}\left(\frac{\lambda_q^3}{g_\rho}\right)^2\ .
\end{equation}
As we will show, the Wilson coefficients contain a suppression factor $\delta_q$, compared with partial compositeness, for each factor $u_R$ or $d_R$ present in the corresponding operator, although in some cases extra suppression factors can also be present (see Eq.~(\ref{C4f1RoCpc})). These suppression factors change the main bounds on $f$ and $m_\rho$, as those arising from the neutron dipole moments and $\epsilon_K$, allowing $f$ and $m_\rho$ even somewhat below the TeV for these processes.

\subsubsection{Dipole operators}
We consider dipole operators involving a Right-handed quark of the first generation, neglecting $\lambda^1_R$ we obtain:
\begin{equation}
C_d^{1_Ra_L} \sim \frac{1}{m_1}\frac{1}{(4\pi)^2}\frac{v}{m_\rho^2g_\rho}\left[g_\rho y\lambda_L^a\lambda_L^3+ y^2\lambda_L^3\lambda_R^3\right]\ . \label{CdaL1R}
\end{equation}
Using Eqs.~(\ref{yjk}), (\ref{lambdaCKM}) and (\ref{mixqR}) one can check that the term of ${\cal O}(y)$ dominates over the term of ${\cal O}(y^2)$. Using these equations, and comparing with the case of anarchic partial compositeness, we obtain:
\begin{equation}
C_d^{1_Ra_L} \sim (C_d^{1_Ra_L})_{\rm pc} \ \delta_q. \label{CdaL1RoCpc}
\end{equation}

The $\Delta F=0$ dipole operator of the down quark of the first generation is strongly constrained by the neutron EDM~\cite{Olive:2016xmw}. In anarchic partial compositeness one obtains a bound $f\gtrsim 5$~TeV~\cite{Panico:2015jxa}. Eq. (\ref{CdaL1RoCpc}) relaxes the bound by a factor $\sqrt{\delta_q}\sim 10^{-1}\lambda_q^3/g_\rho$. A similar result is obtained for the up-quark, with a somewhat less stringent bound.

$\Delta_F=1$ dipole operators with $d_R$ or $u_R$ quarks have the same suppression factor. $C_d^{s_Ld_R}$ is constrained by $\epsilon'/\epsilon$~\cite{Gedalia:2009ws}, leading to $f\gtrsim 1.2$~TeV in anarchic partial compositeness~\cite{Panico:2015jxa}. Eq.~(\ref{CdaL1RoCpc}) relaxes this bound by a factor $\sqrt{\delta_q}$. However the Wilson coefficient of the chiral flipped operator, $C_d^{s_Rd_L}$, does not have this extra suppression in our model, see sec.~\ref{sec-bounds-pc} for a brief discussion on this operator. 

\subsubsection{Penguin operators}
$C_{Z_Q}^{ab}$ generates $\Delta F=1$ flavor violating $Z$-interactions. Neglecting the effect of $\lambda^1_R$, for operators involving $u_R$ or $d_R$ we obtain:
\begin{equation}\label{CZaR1R}
C_{Z_R}^{1a}\sim\frac{g}{c_W}\frac{v^2}{m_\rho^2g_\rho^2}\left[g_\rho y\lambda^3_L\lambda^a_R+y^2(\lambda^3_L)^2\right]  \ . 
\end{equation}
Trading factors of $\lambda$ and $y$ by masses and CKM mixing angles, Eqs.~(\ref{yjk}), (\ref{lambdaCKM}) and (\ref{mixqR}), one can check that the term of ${\cal O}(y^2)$ is suppressed compared with the one of ${\cal O}(y)$. By comparing with the case of anarchic partial compositeness:
\begin{equation}\label{CZ-sd}
C_{Z_R}^{1a}\sim (C_{Z_R}^{1a})_{\rm pc}\ \delta_q \ .
\end{equation}

The most important bound on this class of operators aises from $s-d$ transitions, that are strongly constrained by the decay $K_L\to\mu^+\mu^-$~\cite{Buras:2011ph}. In anarchic partial compositeness without custodial protection of the couplings, $C_{Z_R}^{sd}$ leads to $m_\rho\gtrsim 0.6 g_\rho/\lambda_q^3$~TeV, Eq.~(\ref{CZ-sd}) relaxes this bound by a factor $\sqrt{\delta_q}$. The bounds on $C_{Z_R}^{sd}$ are relaxed if the $Z$ coupling is protected by a discrete symmetry~\cite{Agashe:2006at}.

\subsubsection{Four fermion operators}
The largest source of $\Delta F=2$ flavor violation arises from mixed-chirality four fermion operators. Considering operators with one power of $u_R$ or $d_R$, and taking $\lambda^1_R$ very small, we obtain:
\begin{equation} \label{C4f1R}
C_{4f}^{a_Lb_Lc_R1_R} \sim \frac{1}{m_\rho^2g_\rho^2} \frac{v^2}{m_\rho^2}\lambda^a_L\lambda^b_L[g_\rho y\lambda^c_R\lambda^3_L+y^2(\lambda^3_L)^2+y^2\lambda^c_R\lambda^3_R] \ .
\end{equation}
Using Eqs.~(\ref{yjk}), (\ref{lambdaCKM}) and (\ref{mixqR}), one can check that the term of ${\cal O}(y^2)$ is suppressed compared with the one of ${\cal O}(y)$. By comparing with the case of anarchic partial compositeness:
\begin{equation}\label{C4f1RoCpc}
C_{4f}^{a_Lb_Lc_R1_R}\sim(C_{4f}^{a_Lb_Lc_R1_R})_{\rm pc}\left(\frac{v}{f}\right)^2\ \delta_q \ .
\end{equation}

The most important bounds on $\Delta F=2$ operators arise from the $K^0-\bar K^0$ system, that strongly constrains $C_{4f}^{d_Ls_Ld_Rs_R}$, leading to $m_\rho\gtrsim 10$~TeV in anarchic partial compositeness~\cite{Csaki:2008zd}. Eq.~(\ref{C4f1RoCpc}) relaxes this bound by a factor $\sqrt{\delta_q}v/f$. Other four fermion operators involving $u_R$ or $d_R$ have less stringent bounds.

\subsubsection{Bounds from partial compositeness}\label{sec-bounds-pc}
As discussed in this section, the bounds from flavor violating operators involving $u_R$ or $d_R$ allow $f$ and $m_\rho$ lighter than 1~TeV. However, since the Wilson coefficients of operators that do not involve these quarks are dominated by partial compositeness, the bounds arising from these operators are not relaxed in our scenario, compared with models of anarchic partial compositeness. Following Refs.~\cite{KerenZur:2012fr,Panico:2015jxa}, the main bounds arise from the $\Delta F=2$ operators $(\bar s_R d_L)^2$ and ${\cal Q}_{4f}^{d^i_Ld^j_Ld^i_Ld^j_L}=(\bar d_L^i\gamma^\mu d_L^j)^2$, with $i>j$, that require: $m_\rho\gtrsim 6$~TeV and $m_\rho\gtrsim \lambda_q^3/\lambda_u^3\times 6-7$~TeV, respectively.
Whereas the main $\Delta F=1$ operators are the electromagnetic and chromomagnetic dipoles ${\cal Q}_d^{s_Rd_L}$, that give a bound: $f\gtrsim 1.2$~TeV. Thus in the present scenario these bounds do not change compared with anarchic partial compositeness, and they are the most stringent ones.

There are other operators that can give stringent bounds, as $Q_{Z_L}^{sd}$, however discrete symmetries can protect the $Z$ couplings relaxing the bounds from this operator~\cite{Agashe:2006at}.

\subsection{Leptons}
We consider the effects of flavor and CP violation in the lepton sector, realized as discussed in sec.~\ref{sec-leptons}. We define:
\begin{equation}
\delta_{\ell}\equiv \frac{\sqrt{m_em_\tau}}{v\ g_\rho} \simeq \frac{10^{-4}}{g_\rho} \ .
\end{equation}
The Wilson coefficient of operators involving leptons are suppressed by one power of $\delta_\ell$ for each power of the electron field, either $e_L$ or $e_R$, compared with anarchic partial compositeness. As we will show, due to this suppression the most important problems from flavor and CP violation in the lepton sector can be alleviated, such that $m_\rho\sim$ TeV is compatible with the bounds.

\subsubsection{Dipole operators}
For the $\Delta F=0$ dipole operator of the electron we obtain:
\begin{equation}\label{Cd-e}
C_d^{ee}\sim (C_d^{ee})_{\rm pc}\ \delta{\ell}^2 \ ,
\end{equation}
with two powers of $\delta_\ell$ due to the presence of $e_L$ and $e_R$ in the operator.

The electron EDM gives very strong constraints on the Wilson coefficient $C_d^{ee}~$~\cite{Olive:2016xmw}. In anarchic partial compositeness: $f\gtrsim 38$~TeV~\cite{Panico:2015jxa}. Eq. (\ref{Cd-e}) relaxes this bound by a factor $\delta_l$. 

For $\Delta F=1$ dipole operators involving the first and second generations of charged leptons, $\mu\to e\gamma$, we obtain:
\begin{equation}\label{Cd-mue}
C_d^{\mu e}\sim (C_d^{\mu e})_{\rm pc}\ \delta{\ell} \ .
\end{equation}
In this case just one power of $\delta_\ell$ is present since the operator involves only one of the chiralities of the electron, either $e_L$ or $e_R$. The same suppression factor is present for $\tau\to e\gamma$, under the assumptions of sec.~\ref{sec-leptons}.

$C_d^{\mu e}$ is strongly constrained by $\mu\to e\gamma$~\cite{Olive:2016xmw}. The bound $f\gtrsim 25$~TeV is obtained in anarchic partial compositeness, whereas Eq. (\ref{Cd-mue}) relaxes this bound by a factor $\sqrt{\delta_l}\sim 10^{-2}/\sqrt{g_\rho}$. 

\subsubsection{$\Delta F=1$ penguin operators}
For the flavor violating $Z$-interactions we obtain: 
\begin{equation}\label{CZ-mue}
C_{Z_{L,R}}^{\mu e}\sim (C_{Z_{L,R}}^{\mu e})_{\rm pc}\ \delta{\ell} \ ,
\end{equation}
the same suppression factor is present for the $\tau-e$ coupling.

$C_Z^{\mu e}$ contributes to $\mu-e$ conversion in nuclei~\cite{KerenZur:2012fr}. In anarchic partial compositeness the bound is $m_\rho\gtrsim 2 \sqrt{g_\rho}$~TeV, Eq. (\ref{CZ-mue}) relaxes this bound by a factor $\sqrt{\delta_l}$.

\section{Conclusions}\label{sec-conclusions}
We have considered a theory of composite Higgs with anarchic flavor, taking the SM fermions as elementary fields external to the strongly interacting sector. We have introduced linear and bilinear interactions at a UV scale much larger than the scale of resonances, that has been taken as $m_\rho\sim$ few TeV. We have considered the case in which, under some suitable assumptions for the energy evolution of the couplings, the interplay of both interactions can generate the hierarchy of fermion masses and the pattern of mixing angles in the sector of quarks and leptons. In particular, we have studied the case where the masses of the fermions of the first generation, being all of the same order, are dominated by bilinear interactions, whereas the masses of the fermions of the second and third generations are dominated by partial compositeness. In the sector of quarks, this scenario can be realized if, at the scale $m_\rho$, the bilinear coupling is of order $y_{u,d}^{\rm SM}$ and the linear interactions of $u_R$ and $d_R$ are very small. On the other hand, a non-vanishing Cabibbo angle requires non-vanishing linear interaction of $(u_L,d_L)$. For the leptons, if the neutrinos are Majorana particles, neutrino masses can be generated by the dimension five Weinberg operator with an anarchic coupling at high energy scales, naturally leading to large mixing angles and neutrino masses of the same order. Being this the case, the linear interactions of $e_L$ and $e_R$ can be taken very small at the scale $m_\rho$, and the electron mass is generated by the bilinear interaction. Considering hierarchical linear interactions for both chiralities of $\mu$ and $\tau$, the diagonalization matrices of the charged leptons have small mixing angles.

We have presented the contributions to the main flavor violating operators from the linear and the bilinear interactions. We have shown that, in the scenario described in the previous paragraph, some of the most stringent bounds can be relaxed, as those arising from the neutron dipole moments, those from chiral-mixed operators in the system $K^0-\bar K^0$, the electron EDM and $\mu-e$ processes. In anarchic partial compositeness, these processes require $m_\rho$ and $f$ larger than $\sim 10$ and $\sim 5$ TeV, respectively, in the quark sector, and $f\gtrsim 20-40$~TeV in the lepton sector. In the scenario proposed in this paper, these bounds are compatibles with $m_\rho$ and $f$ below the TeV.

Flavor and CP violating processes that do not involve $u_R$ or $d_R$, do not change compared with the usual anarchic partial compositeness. Some of these processes provide the most stringent bounds in the model proposed, pushing the scales $m_\rho$ and $f$ to $\sim 6-7$ and $\sim 1$~TeV, respectively~\cite{KerenZur:2012fr,Panico:2015jxa}. In our scenario, effects of flavor violation should be observed first in processes involving the operators $\bar s_RF_{\mu\nu}\sigma^{\mu\nu}d_L$, $(\bar s_R d_L)^2$, $(\bar d_L^i\gamma^\mu d_L^j)^2$, with $i>j$. These operators give contributions to the Kaon and $B$-meson systems, as well as $K\to 2\pi$.

The phenomenology of the model at colliders is similar to the case of anarchic partial compositeness. The most important difference is that resonances associated to Right-handed fermions of the first generation, as well as the Left-handed electron, having tiny mixing with  these fermions, will be difficult to excite.

Although in the present model the bounds on $m_\rho$ from flavor violation are alleviated compared with anarchic partial compositeness, the flavor violating operators mentioned above require $m_\rho\gtrsim 6-7$~TeV. This bound introduces some tension with naturalness, that requires $m_\rho$ and $f$ around the TeV, and puts the resonances out of reach of direct production at the LHC. Finding a way to suppress the contributions to those operators would relax this tension.

\section*{Acknowledgements}
We thank Andrea Wulzer for very interesting discussions on the beginning of this work, and Michele Redi for clarifications concerning dipole operators with a pNGB Higgs. We also thank the University of Sussex for hospitality in the initial stages of this project. This work was partially supported by the Argentinian ANPCyT PICT 2013-2266.

\appendix
\section{A minimal model}\label{ap-minimal-model}
An explicit calculation of the Wilson coefficients of the operators of Eqs.~(\ref{operatorQdp})-(\ref{operatorQZ}) can be done by considering an effective description of the SCFT. In this section we consider a theory with two sites: one site called elementary and containing the same fields and interactions as the SM, except for the Higgs field, and another site containing the Higgs and the first level of resonances of the composite sector~\cite{Contino:2006nn}. Spin 1 resonances can be obtained by gauging a symmetry in the second site, for simplicity we consider an SU(3)$_c\times$SU(2)$_L\times$U(1)$_Y$ group, custodial symmetry can be included straightforwardly. Fermionic resonances can be described by adding Dirac fermions in the second site, we add one vector-like fermion for each SM quark and we assign them to the same representations as the SM fermions. Both sites are connected by non-linear sigma model fields that parametrize the breaking of the elementary and composite symmetries to the diagonal subgroup, providing masses to the gauge fields associated to the broken generators, and leaving a set of massless vector fields associated to the unbroken generators. The non-linear sigma model fields allow to introduce the usual linear interactions between the elementary and composite fermions. We also add Yukawa interactions bilinear in the elementary fermions, as well as Yukawa interactions bilinear in the composite fermions. The couplings of the composite sector, collectively called $g_\rho$, are assumed to be larger than the couplings of the elementary one but in the perturbative regime: $g_{el}\ll g_\rho\ll 4\pi$. The SM gauge couplings are given by: $g_{SM}^{-2}=g_\rho^{-2}+g_{el}^{-2}$. We will use small letters to denote the elementary fields and capital letters to denote the composite ones. Below we discuss the sector of quarks, for the leptons we will not show the calculations, but the results presented in the previous sections have been obtained by following the same strategy.

The Lagrangian can be written as:
\begin{align}\label{l1}
{\cal L}={\cal L}_{el}+{\cal L}_{mix}+{\cal L}_{cp}\ ,
\end{align}
where ${\cal L}_{el}$ is the Lagrangian of the elementary sector, containing the kinetic terms of the gauge and fermion fields of the first site, that coincide with the corresponding terms of the SM. ${\cal L}_{mix}$ contains the terms mixing both sites and ${\cal L}_{cp}$ the terms of the second site.

${\cal L}_{cp}$ contains the following terms: 
\begin{align}\label{lfermioncptoy}
{\cal L}_{cp}\supset & \bar Q (i\Dslash -m_Q) Q+\bar D (i\Dslash-m_D)D+y_\rho\bar Q_LHD_R+\hat y_\rho\bar Q_R HD_L+{\rm h.c.}
\end{align}
$Q=(Q^u,Q^d)$ and $D$ denote the multiplets of fermionic resonances, with generation indices being implicit. We have shown the terms involving fields of the down-sector only, the extension to the up-sector is straightforward once an SU(2)$_L$ singlet $U$ is added. Leptons can also be included by adding vector-like fermions with the same quantum numbers as the SM fermions. As discussed before, there are also kinetic terms for the gauge fields of the second site, as well as kinetic terms and a potential for the Higgs.

By working in the gauge where the non-linear sigma model fields are removed:
\begin{equation}\label{lfermionmixtoy}
{\cal L}_{mix}\supset \frac{f^2}{4}(g_{el}a_\mu^a-g_\rho A_\mu^a)^2+f\ (\lambda_q\bar q_L Q+\lambda_d\bar d_RD)+y\bar q_LHd_R +{\rm h.c.}
\end{equation}
The fields $a_\mu$ and $A_\mu$ denote the gauge fields of the elementary and composite gauge groups, respectively. 
Also in this case it is straightforward to complete the up-sector, that is not shown for simplicity.  

Since $g_\rho\gg g_{el}$, the massive spin one states are mostly composite. In the limit of vanishing elementary couplings, these states are fully composite, with mass: $m_\rho=g_\rho f/\sqrt{2}$. 

As discussed in sec.~\ref{sec-scenario}, we choose a basis where $\lambda$ is diagonal, whereas the Yukawa couplings $y$ and $y_\rho$, as well as the masses of the composite fermions: $m_Q$ and $m_D$, are anarchic.~\footnote{Notice that all the eigenvalues of $m_Q$ and $m_D$ are of the same order, thus there are no hierarchies between these states. After the linear mixing is taken into account, the masses of the composite fermions are shifted, as usual in partial compositeness, the states with large mixing obtaining a larger contribution than those with small mixing.} These masses are assumed to be of order $m_\rho$, and the Yukawa couplings of the composite site are taken as: $y_\rho,\hat y_\rho\sim g_\rho$.

In the basis: $\Psi_L=(q^d_L,Q^d_L,D_L)$, $\Psi_R=(d_R,Q^d_R,D_R)$, the mass matrix is given by:
\begin{equation}\label{Mminimal}
M=\left(\begin{array}{ccc}yv/\sqrt{2}&f\lambda_q&0\\0&m_Q&y_\rho v/\sqrt{2}\\f\lambda_d&\hat y_\rho v/\sqrt{2}&m_D\end{array}\right) \ ,
\end{equation}
and the corresponding Yukawa by: $Y=\frac{d}{dv}M$.

\subsection{Mass basis}\label{sec-mass-basis}
To obtain the Wilson coefficients (\ref{Cd})-(\ref{C4f}) we rotate to the mass basis. We consider first the case of one generation and compute the lightest eigenvalue of the mass matrix and the corresponding eigenvectors. To ${\cal O}\left(\frac{v}{f}\right)$, the lightest mass and eigenvectors are:
\begin{align}\label{eigensystem1}
&m^{(1)}\simeq \frac{v}{\sqrt{2}}\frac{|m_Q m_D y+y_\rho f^2\lambda_q\lambda_d|}{(m_Q^2+f^2\lambda_q^2)^{1/2}(m_D^2+f^2\lambda_d^2)^{1/2}}
\ ,
\\
&v_L^{(1)}\simeq \left(\frac{m_Q}{(m_Q^2+f^2\lambda_q^2)^{1/2}},-\frac{f\lambda_q}{(m_Q^2+f^2\lambda_q^2)^{1/2}},\frac{v}{\sqrt{2}}\frac{-y\lambda_dfm_Q+y_\rho\lambda_qfm_D}{(m_Q^2+f^2\lambda_q^2)^{1/2}(m_D^2+f^2\lambda_d^2)}\right)
\ ,\label{eigensystem2}
\\
&v_R^{(1)}\simeq \left(\frac{m_D}{(m_D^2+f^2\lambda_d^2)^{1/2}},\frac{v}{\sqrt{2}}\frac{y_\rho\lambda_dfm_Q+y\lambda_qfm_D}{(m_Q^2+f^2\lambda_q^2)(m_D^2+f^2\lambda_d^2)^{1/2}},-\frac{f\lambda_d}{(m_D^2+f^2\lambda_d^2)^{1/2}}\right)
\ .\label{eigensystem3}
\end{align}

In the scenario proposed in sec.~\ref{sec-scenario}, for the first generation the first term in the numerator of Eq.~(\ref{eigensystem1}) dominates, leading to $m_d\sim y v/\sqrt{2}$, whereas for the other generations the second term dominates: $m_{j}\sim \lambda_q^j\lambda_d^jv/g_\rho\sqrt{2}$, with $j=s,b$. For these estimates we have used: $m_{Q,D}\sim g_\rho f$.

In the following we will use a hat for the matrices in the basis of mass eigenstates, we will order this basis by increasing mass eigenvalue, with $n=1$ for the lightest sate that is associated with the SM down-type quarks, and $n>1$ for the heavier beyond the SM states. We will use subindices for the matrix elements of the operators in this basis. 

\subsection{Flavor and CP violation}\label{subsec-fcpv1}
In this section we compute the Wilson coefficients (\ref{Cd})-(\ref{C4f}) with just one flavor. We will only show the results for the down-type fermions of the model, the results for the up sector and charged leptons are very similar and can be computed straightforwardly.

We start with $C_d$, we estimate its size by computing the 1-loop contribution with the physical Higgs running in the loop, Fig.~\ref{fig-dipole}. Following Ref.~\cite{Agashe:2008uz}, the main contribution is proportional to
\begin{equation}
A=\sum_{n>1}\frac{\hat Y_{1n}\hat Y_{n1}}{m^{(n)}}
\end{equation}
with $\hat Y_{\ell n}$ the Yukawa coupling in the mass basis, and where we have assumed that $m^{(n)}\gg m_h$. A simple algebraic manipulation allows to simplify the calculation by noticing that the following equation requires just the eigenvalue and eigenvectors of the lightest mode of the mass matrix $M$:
\begin{equation}\label{trick-chdi}
A=(U_{L}^\dagger Y M^{-1}Y U_{R})_{11}-\frac{1}{m^{(1)}}(U_{L}^\dagger Y U_{R})_{11}^2 \ , 
\end{equation}
where $U_{L,R}$ are the rotation matrices that diagonalize $M$. Since $(U_{L})_{n1}=v_L^{(1)}$, $(U_{R})_{n1}=v_R^{(1)}$, it is not necessary to perform the full diagonalization, that requires the calculation of all the eigenvalues and eigenvectors of $M$. By using Eqs.~(\ref{Mminimal})-(\ref{eigensystem3}) in (\ref{trick-chdi}), we obtain:
\begin{align}
A\simeq vf^2&\left[\lambda_d^2\left(\frac{yy_\rho^2}{m_Q^2m_D^2}+3\frac{yy_\rho\hat y_\rho}{m_Qm_D^3} \right)+\lambda_q^2\left(\frac{yy_\rho^2}{m_Q^2m_D^2}+3\frac{yy_\rho\hat y_\rho}{m_Q^3m_D} \right)\right.
\\
&\left.-\lambda_d\lambda_q\left(3\frac{y^2\hat y_\rho}{m_Q^2m_D^2}+2\frac{y^2y_\rho}{m_Qm_D^3}+2\frac{y^2y_\rho}{m_Q^3m_D}+3\frac{y_\rho^2\hat y_\rho}{m_Q^2m_D^2}\right)\right] \ ,
\end{align}
that is in agreement with the estimate of Eq.~(\ref{Cd}).

$C_Z$ can be obtained by computing the matrices of $Z$-couplings: $G_{L,R}^Z$, in the original basis, and then projecting them with the eigenvectors of the lightest modes. Subtracting the SM term and expanding in powers of $\lambda$ and $y$ we obtain:
\begin{eqnarray}
&(\hat G_R^Z)_{11}\simeq &\frac{g}{c_W}\frac{v^2f^2}{m_D^2}\left(\frac{y_\rho^2\lambda_d^2}{2m_D^2}+\frac{y^2\lambda_q^2}{2m_Q^2}-\frac{yy_\rho\lambda_d\lambda_q}{m_Dm_Q}\right)
\ , \label{g1RZ}
\\
&(\hat G_L^Z)_{11}\simeq &-(\hat G_R^Z)_{11} (q\leftrightarrow d,Q\leftrightarrow D)
\ , \label{g1LZ} 
\end{eqnarray}
in agreement with the estimate of Eq.~(\ref{CZ}).

$C_{4f}$ is induced by exchange of a gluon resonance. Going to the mass basis, the mixing between elementary and composite fermions generates interactions between the light fermions and the massive gluon. The couplings of Left- and Right-handed light fermions are:
\begin{eqnarray}
&(\hat g_R)_{11}\simeq &g_\rho\frac{f^2\lambda_d^2}{m_D^2}\left[1+\frac{v^2}{2}\left(2\frac{y^2}{m_D^2}+\frac{y_\rho^2}{m_Q^2}+2\frac{y_\rho\hat y_\rho}{m_Dm_Q}\right)\right] \nonumber
\\
&&+ g_\rho\frac{v^2}{2}\left[\lambda_q^2\frac{y^2f^2}{m_Q^4}-2\frac{yf^2\lambda_d\lambda_q}{m_Dm_Q}\left(\frac{y_\rho}{m_D^2}+\frac{y_\rho}{m_Q^2}+\frac{\hat y_\rho}{m_Dm_Q}\right)\right] 
\ , \label{g1RG}
\\
&(\hat g_L)_{11}\simeq &(\hat g_R)_{11} (q\leftrightarrow d,Q\leftrightarrow D)
\ , \label{g1LG}
\end{eqnarray}
where we have expanded to second order in powers of $\lambda$ and $v/f$. Matching $C_{4f}=(\hat g)_{11}^2/m_{\rho}^2$, with the corresponding chirality subindices, and expanding in powers of $y$ to second order, leads to the estimate of Eq.~(\ref{C4f}).

\section{MCHM}\label{ap-MCHM5}
In this section we will consider the case in which the Higgs is a pNGB generated by the spontaneous breaking of a global symmetry of the SCFT. We will study the scenario with global symmetry SO(5)$\times$U(1)$_X$, with the Higgs arising from SO(5)/SO(4), known as the Minimal Composite Higgs Model (MCHM)~\cite{Agashe:2004rs}. An SU(2)$_L\times$U(1)$_Y$ subgroup is gauged by the EW fields of the SM. The SCFT also has a global SU(3), that is gauged by the color interactions of QCD. As described in sec.~\ref{sec-scenario}, the SM fermions are taken as elementary fields. We will consider an effective description of the SCFT in which we will only include one level of fermionic resonances. We will assume that there are three generations of composite fermions. As mentioned in sec.~\ref{sec-flavor-cp-v}, we will choose SO(5) representations for fermions where, to leading order, the Yukawa coupling is aligned with the mass for the would be zero modes corresponding to the SM fermions~\cite{Agashe:2009di}. Interestingly, this scenario contains a new flavor structure, the $d$-term associated to the Nambu-Goldstone boson (NGB) nature of the Higgs, that provides new sources of flavor violation compared with the minimal model of Ap.~\ref{ap-minimal-model}.

The Higgs is described by the field $\Pi=\Pi_{\hat a}T^{\hat a}$, with $T^{\hat a}$ the broken generators of SO(5), $\hat a=1,\dots 4$. The fundamental object to build the effective theory is the NGB matrix
\begin{equation}
U=e^{i\sqrt{2}\Pi/f} \ .
\end{equation}
By following the standard Callan-Coleman-Wess-Zumino procedure, one can build the operators $d_\mu$ and $e_\mu$, defined by: $U^\dagger D_\mu U=e^a_\mu T^a+d^{\hat a}_\mu T^{\hat a}$, $T^a$ being the unbroken generators and $D_\mu$ the usual covariant derivative. 

The composite fermions are assumed to transform with linear irreducible representations of the unbroken group SO(4), and with non-linear representations of SO(5). We will use capital letters to label SO(5) representations and small letters for SO(4) representations, such that a given $\bR$ of SO(5) decomposes under SO(4) as: $\bR\sim\oplus_j \br_j$. To leading order the effective Lagrangian of the composite sector is 
\begin{align}\label{LcpNGB}
{\cal L}_{cp}\supset & \frac{f^2}{4}d_\mu^{\hat a}d^{\hat a \mu}+\sum_\br \left[i\bar\Psi_\br (\partialslash+i\eslash-ig_x \Xslash) \Psi_\br+\bar\Psi_\br m_\br \Psi_\br\right] \nonumber \\
&+\sum_{\br,\bs}\left[ic_{\br\bs}\bar\Psi_\br \dslash \Psi_\bs+{\rm h.c.}+ \frac{C_{\br\bs}}{f^2}(\bar\Psi_\br\Gamma\Psi_\br)(\bar\Psi_\bs\Gamma\Psi_\bs)+ \frac{\tilde C_{\br\bs}}{f^2}(\bar\Psi_\br\Gamma\Psi_\bs)(\bar\Psi_\bs\Gamma\Psi_\br)\right] \ .
\end{align}
The first line contains the kinetic term of the NGB fields and fermions, flavor and generation indices are understood. The first term of the second line contains interactions between the NGB and composite fermions in different representations of SO(4). The coupling $c_{\br\bs}$ can distinguish chiralities, we will allow different couplings for the Left- and Right-handed fermions in our analysis. The 4-fermion interactions can be generated, for example, by exchange of resonances. $\Gamma$ is a generic matrix, in Dirac and color space, that depends on the quantum numbers of the resonance exchanged. Particularly interesting is the case of $\Gamma=\gamma^\mu P_{L,R}T^\alpha$, with $T^\alpha$ the color generators, that, after integration of the fermionic resonances, contributes to the Wilson coefficient $C_4$ of the mixed chirality four-fermion operator.

The size of the coefficients of ${\cal L}_{cp}$ can be estimated by using suitable power-counting rules~\cite{Giudice:2007fh,DeSimone:2012fs,Grojean:2013qca}, leading to $c_{\br\bs}$, $C_{\br\bs}$ and $\tilde C_{\br\bs}$ of ${\cal O}(1)$. Anarchy of the SCFT implies that all the elements of the matrices $m_\br$, $c_{\br\bs}$, $C_{\br\bs}$ and $\tilde C_{\br\bs}$, in flavor space, are of the same order.

Although we have not included vector resonances in ${\cal L}_{cp}$, they can be added. In the case of an effective description in a two-site model, this can be done by gauging the global symmetries of the second site, as in Ap.~\ref{ap-minimal-model}~\cite{Contino:2011np,Panico:2011pw,Azatov:2013ura,Panico:2015jxa}. In the low energy theory these resonances can be integrated-out, giving contributions similar to those of Eq.~(\ref{LcpNGB}). 

The Lagrangian of the elementary sector, ${\cal L}_{el}$, contains the usual kinetic terms for fermions and gauge bosons. As usually done in the literature~\cite{Agashe:2004rs}, we find it useful to add spurious degrees of freedom in the elementary sector to embed the corresponding fields into irreducible representations of SO(5). We will call the SO(5) representations of the elementary quarks $\bR_q$, $\bR_u$ and $\bR_d$, choosing the same representations for the three generations. These representations can be decomposed in irreducible representations of SO(4), we will call $\br_q$ the SO(4) representation of $\bR_q$ that contains the SM doublet $q_L$, and $\br_{u,d}$ the ones containing the singlets. Similarly, the elementary gauge fields will be embedded into the adjoint representation of SO(5)$\times$U(1)$_X$, with $Y=T^{3R}+X$.

Besides ${\cal L}_{cp}$ and ${\cal L}_{el}$, there is also a mixing Lagrangian given by:
\begin{align}\label{LmixNGB}
{\cal L}_{mix}\supset & f\sum_\br\left[(\bar q_LU)_\br\lambda_{q{\br}} \Psi_\br+(\bar d_RU)_\br\lambda_{d{\br}} \Psi_\br+(\bar q_LU)_\br y_d^\br(U^\dagger d_R)_\br\right]+{\rm h.c.}
\end{align}
where we have not included the terms corresponding to the up-sector for simplicity, but it is straightforward to add them. $\br$ is any irreducible representation of SO(4) contained in $\bR_q$ or $\bR_d$. $\phi_\br$ is the projection of $\phi$, that transforms with an irreducible representation of SO(5), into $\br$. The product of two representations $\br$ is projected into the SO(4) singlet.
 
The interactions between the elementary and composite sectors explicitly break the SO(5) global symmetry, inducing a potential for $\Pi$ at loop-level. This potential can misalign the vacuum and trigger EWSB.
A general vacuum can be characterized by the variable $\epsilon=\sin(v/f)$, with $v=\langle h\rangle$ and $h=\Pi^{\hat 4}$ being the physical Higgs~\cite{Agashe:2004rs}. 

\subsection{MCHM$_5$}\label{MCHM5}
As an example we consider the down sector of the MCHM$_5$, we embed $q_L$ and $d_R$ in the same representation: $\bR_q=\bR_d={\bf 5}_{-1/3}\sim ({\bf 2},{\bf 2})_{-1/3}\oplus({\bf 1},{\bf 1})_{-1/3}$. In this case: $\br_q=({\bf 2},{\bf 2})$ and $\br_d=({\bf 1},{\bf 1})$. For the composite fermions we consider: $\Psi_{\bf 4}\sim({\bf 2},{\bf 2})_{-1/3}$ and $\Psi_{\bf 1}\sim({\bf 1},{\bf 1})_{-1/3}$:
\begin{equation}
\Psi_{\bf 4}=\frac{1}{\sqrt{2}}\left(\begin{array}{c}i(V-U)\\V+U\\i(D-D')\\D+D'\end{array}\right) \ , \qquad \Psi_{\bf 1}= \tilde D \ ,
\end{equation}
where $U$ and $V$ have electric charge $2/3$ and $-4/3$, respectively, whereas the other states have $Q=-1/3$. The charges of the down-type resonances under SU(2)$_L$ and SU(2)$_R$, in units of $1/2$, are: $D^{(-,+)}$, $D^{'(+,-)}$ and $\tilde D^{(0,0)}$. 

A description of the up-sector requires the presence of new fermionic resonances, for example a $\Psi'_{\bf 4}\sim({\bf 2},{\bf 2})_{2/3}$ and $\Psi'_{\bf 1}\sim({\bf 1},{\bf 1})_{2/3}$. It is straightforward to include these states in the model, for simplicity we will not consider them in the analysis of this section. Similarly, leptons can be included by choosing appropriate representations, as $\bR_\ell=\bR_e={\bf 5}_{-1}$, with $\br_\ell=({\bf 2},{\bf 2})$ and $\br_e=({\bf 1},{\bf 1})$, and the addition of $\Psi^\ell_{\bf 4}\sim({\bf 2},{\bf 2})_{-1}$ and $\Psi^e_{\bf 1}\sim({\bf 1},{\bf 1})_{-1}$. In this case the sector of charged leptons is similar to the sector of down-type quarks. The mass matrix, its eigenvectors and eigenvalues can be obtained from the ones of the down-type quark, by changing $q\to\ell$ and $d\to e$. The matrices of couplings with the gauge bosons can be computed straightforwardly.

\subsection{Interactions and mass basis}
Similarly to Ap.~\ref{ap-minimal-model}, to estimate the size of the Wilson coefficients, we find it useful to work in the mass basis. We consider first the case of one generation and compute the lightest eigenvalue and eigenvectors of the mass matrix. In the basis $\Psi_L=(d_L,D_L,D'_L,\tilde D_L)$, $\Psi_R=(d_R,D_R,D'_R,\tilde D_R)$, the mass matrix of the down-sector is given by:
\begin{equation}
M=\left(\begin{array}{cccc}
\frac{\epsilon\sqrt{1-\epsilon^2}}{\sqrt{2}}f(y_{d\bf4}-y_{d\bf1}) & \frac{1+\sqrt{1-\epsilon^2}}{2}f\lambda_{q{\bf4}} & \frac{\sqrt{1-\epsilon^2}-1}{2}f\lambda_{q{\bf4}} & -\frac{\epsilon}{\sqrt{2}}f\lambda_{q{\bf1}} \\
\frac{\epsilon}{\sqrt{2}}f\lambda_{d{\bf4}} & m_{d\bf4} & 0 & 0 \\
\frac{\epsilon}{\sqrt{2}}f\lambda_{d{\bf4}} & 0 & m_{d\bf4} & 0 \\
\sqrt{1-\epsilon^2}f\lambda_{d{\bf1}} & 0 & 0 & m_{d\bf1} 
\end{array}\right) \ .
\end{equation}
For the calculation of different Wilson coefficients, the couplings with the Higgs and the EW gauge bosons are required:
\begin{equation}
{\cal L}_{\rm int}\supset h\bar\Psi_LY\Psi_R+{\rm h.c.}+\bar\Psi_L(G^Z_LZ_\mu+G^\ddh_L\partial_\mu h)\gamma^\mu\Psi_L+(L\to R) \ .
\end{equation}
where $Y=\frac{\partial M}{\partial v}$, whereas $G^{\ddh}$ and $G^Z$ are the matrices of couplings with the derivative of the Higgs and with the $Z$, respectively. In the interaction basis:
\begin{align}
&G^Z_L=\frac{g}{c_W}
\left(
\begin{array}{cccc}
 \frac{s_W^2}{3}-\frac{1}{2} & 0 & 0 & 0 \\
 0 & \frac{s_W^2}{3}-\frac{\sqrt{1-\epsilon^2}}{2} & 0 & -\frac{1}{2} c_L \epsilon \\
 0 & 0 & \frac{\sqrt{1-\epsilon^2}}{2}+\frac{s_W^2}{3} & \frac{c_L \epsilon}{2} \\
 0 & \frac{c_L \epsilon}{2} & -\frac{1}{2} c_L \epsilon & \frac{s_W^2}{3} \\
\end{array}
\right) \ , \nonumber
\\
&G^Z_R=\frac{g}{c_W}
\left(
\begin{array}{cccc}
 \frac{s_W^2}{3} & 0 & 0 & 0 \\
 0 & \frac{s_W^2}{3}-\frac{\sqrt{1-\epsilon^2}}{2} & 0 & -\frac{1}{2} c_R \epsilon \\
 0 & 0 & \frac{s_W^2}{3}+\frac{\sqrt{1-\epsilon^2}}{2} & \frac{c_R \epsilon}{2} \\
 0 & \frac{c_R \epsilon}{2} & -\frac{1}{2} c_R \epsilon & \frac{s_W^2}{3} \\
\end{array}
\right) 
\ ,\nonumber
\\
&G^\ddh_L=i\frac{c_L}{f}
\left(
\begin{array}{cccc}
 0 & 0 & 0 & 0 \\
 0 & 0 & 0 & 1 \\
 0 & 0 & 0 & 1 \\
 0 & -1 & -1 & 0 \\
\end{array}
\right)
\ ,
\qquad G^\ddh_R=G^\ddh_L(c_L\to c_R) \ .
\end{align}
As in the previous example, we will use a hat for the matrices in the basis of mass eigenstates: $\hat Y, \hat G^{\ddh}, \hat G^Z$.

We have diagonalized the down-sector performing a perturbative expansion in powers of $v/f$, to ${\cal O}(v^3/f^3)$, below we show our results to ${\cal O}(v/f)$, it is also straightforward to diagonalize the system expanding in powers of $\epsilon$. The lightest mass and eigenvectors are:
\begin{align}\label{eigensystemNGB1}
&m^{(1)}\simeq \frac{v}{\sqrt{2}}\frac{f\lambda_{q{\bf4}}\lambda_{d{\bf4}}m_{d\bf1}-f\lambda_{q{\bf1}}\lambda_{d{\bf1}}m_{d\bf4}+m_{d\bf1}m_{d\bf4}(y_{d\bf1}-y_{d\bf4})}{(m_{d\bf1}^2+f^2\lambda_{q\bf1}^2)^{1/2}(m_{d\bf4}^2+f^2\lambda_{d\bf4}^2)^{1/2}}
\ ,
\\
&v_L^{(1)}\simeq \left(\frac{m_{d\bf4}}{(m_{d\bf4}^2+f^2\lambda_{d\bf4}^2)^{1/2}},-\frac{f\lambda_{q{\bf4}}}{(m_{d\bf4}^2+f^2\lambda_{d\bf4}^2)^{1/2}},0,\frac{v}{\sqrt{2}}\frac{\lambda_{q{\bf1}}m_{d\bf4}m_{d\bf1}+f^2\lambda_{d{\bf1}}\lambda_{d{\bf4}}\lambda_{q{\bf4}}+f\lambda_{d{\bf1}}m_{d\bf4}(y_{d\bf1}-y_{d\bf4})}{(m_{d\bf1}^2+f^2\lambda_{q\bf1}^2)(m_{d\bf4}^2+f^2\lambda_{d\bf4}^2)^{1/2}}\right)
\ ,\label{eigensystemNGB2}
\\
&v_R^{(1)}\simeq \left(\frac{m_{d\bf1}}{(m_{d\bf1}^2+f^2\lambda_{d\bf1}^2)^{1/2}},-\frac{v}{\sqrt{2}}\frac{\lambda_{d{\bf4}}m_{d\bf4}m_{d\bf1}+f^2\lambda_{d{\bf1}}\lambda_{q{\bf1}}\lambda_{q{\bf4}}+f\lambda_{q{\bf4}}m_{d\bf1}(y_{d\bf4}-y_{d\bf1})}{(m_{d\bf4}^2+f^2\lambda_{q\bf4}^2)(m_{d\bf1}^2+f^2\lambda_{d\bf1}^2)^{1/2}},\right. \nonumber\\ & \ \ \ \ \ \ \ \ \ \ \left. -\frac{v}{\sqrt{2}}\frac{\lambda_{d{\bf4}}m_{d\bf1}}{m_{d\bf4}(m_{d\bf1}^2+f^2\lambda_{d\bf1}^2)^{1/2}},-\frac{f\lambda_{d{\bf1}}}{(m_{d\bf1}^2+f^2\lambda_{d\bf1}^2)^{1/2}}\right)
\ .\label{eigensystemNGB3}
\end{align}
As expected there are two contributions to the mass, one from the bilinear coupling, of order $\sim v (y_{d\bf1}-y_{d\bf4})/\sqrt{2}$ and another one involving the linear mixing of order $\sim v(\lambda_{q{\bf4}}\lambda_{d{\bf4}}-\lambda_{q{\bf1}}\lambda_{d{\bf1}})/(g_\rho\sqrt{2})$. As in the case of the non-NGB Higgs, by taking $d_R$ elementary, that corresponds to $\lambda_d^{\bf1,4}\to 0$ for the first generation, the down quark mass is generated by the bilinear interaction. For the second and third generations partial compositeness dominates.

\subsection{Flavor and CP violation}\label{subsec-fcpv2}
As for the minimal model, in this section we show the Wilson coefficients (\ref{Cd})-(\ref{C4f}) with just one flavor and for the down-type fermions. The up sector and charged leptons can be analysed doing similar calculations, for the later an expansion to second order in $y_{\br}$ is required in some cases.

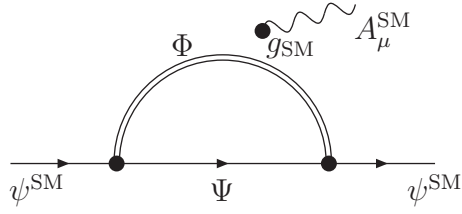
\begin{figure}
\centering
\begin{picture}(170,100)(0,0)
    \ArrowLine(0,40)(40,40)
    \Vertex(40,40){3}
    \Vertex(120,40){3}
    \ArrowLine(40,40)(120,40)
    \ArrowLine(120,40)(160,40)
    \CArc(80,40)(39,0,180)
    \CArc(80,40)(41,0,180)
    \Photon(95,90)(130,100){3}{3}
    \Vertex(95,90){3}
    \Text(10,30)[c]{$\psi^{\rm SM}$}
    \Text(80,30)[c]{$\Psi$}
    \Text(106,84)[c]{$g_{\rm SM}$}
    \Text(130,90)[l]{$A_\mu^{\rm SM}$}
    \Text(150,30)[l]{$\psi^{\rm SM}$}
    \Text(65,85)[c]{$\Phi$}
\end{picture}
\caption{One loop contribution to the dipole operator of the SM fermions. The external $A$ field can be inserted in the fermionic or bosonic propagator. $\Psi$ is a heavy fermion, $\Phi$ can be either a Higgs or a gauge boson. The couplings in the vertices can be gauge, Yukawa or $d-$term couplings.}
\label{fig-dipole}
\end{figure}

Let us start with the dipole operators. There are new contributions to $C_d$ in this model, compared with the minimal one: the $d$-term gives new Feynman diagrams at one loop. For our calculation we follow Ref.~\cite{Antipin:2014mda}.
In Fig.~\ref{fig-dipole} we show the one loop Feynman diagram of the dipole, with exchange of a virtual state $\Phi$. For $\Phi$ being the Higgs boson, there are two possibilities for the vertices: the Yukawa couplings and the derivative interactions arising from the $d-$terms. There are three kind of diagrams in this case, proportional to: $Y^2$, $YG^{\partial h}$ and $(G^{\partial h})^2$. For $\Phi$ being the $Z$ and $W$ the coupling arises from the gauge interactions.

Ref.~\cite{Antipin:2014mda} has provided the corresponding expressions for each kind of diagram in the mass basis. The authors have shown that the contributions from $W^+$ and $W^-$ cancel each other. Below we describe the contributions mediated by the $Z$ and the Higgs with different number of derivatives. Assuming that the heavy fermions are much heavier than the SM particles: $m_\psi\gg m_{q,Z,h}$, the results are simplified as:
\begin{align}\label{eq-aproxloop}
& C_{d}^{Z}\simeq \frac{g_{\rm SM}}{4(4\pi)^2 m^{(0)}}\frac{1}{m_Z^2}\sum_{j>1}(\hat G_L^Z)_{1j} m^{(j)}(\hat G_R^{Z})_{j1} \ , \nonumber\\
& C_{d}^{h}\simeq \frac{g_{\rm SM}}{4(4\pi)^2 m^{(0)}}\sum_{j>1}\frac{(\hat Y)_{1j}(\hat Y)_{j1}}{m^{(j)}} \ , \nonumber\\
& C_{d}^{h\ddh}\simeq -\frac{g_{\rm SM}}{12(4\pi)^2 m^{(0)}}\sum_{j>1}\left[(\hat G_L^\ddh)_{1j}(\hat Y)_{j1}-(\hat Y)_{1j}(\hat G_R^{\ddh})_{j1}\right] \ , \nonumber\\
& C_{d}^{(\ddh)^2}\simeq -\frac{g_{\rm SM}}{8(4\pi)^2 m^{(0)}}\sum_{j>1}(\hat G_L^\ddh)_{1j}m^{(j)}(\hat G_R^{\ddh})_{j1} \ .
\end{align}
By simple algebraic manipulations these equations can be computed in the interaction basis, and then projected to the lightest state by using the eigenvectors, as in the minimal model of Ap.~\ref{ap-minimal-model}. Expanding to second order in $\lambda$ and $y_{\br}$:
\begin{align}
& C_{d}^{Z}\simeq L \frac{v}{\sqrt{2}}\left[-
\frac{\lambda_{q\bf4}\lambda_{d\bf4}}{f\ m_{\bf4}}+\sqrt{2}c_R^\dagger\frac{\lambda_{q\bf4}\lambda_{d\bf1}}{f\ m_{\bf1}}+\sqrt{2}c_L^\dagger\frac{\lambda_{q\bf1}\lambda_{d\bf4}}{f\ m_{\bf1}}-2c_L^\dagger c_R^\dagger\frac{\lambda_{q\bf1}\lambda_{d\bf1}m_{\bf4}}{f\ m_{\bf1}^2}\right. \nonumber\\
&\ \ \ \ \ \ \ \ \ \ \ \ \ \ \ \left.+(y_{\bf4}-y_{\bf1})\left(\frac{2\lambda_{d\bf4}^2-\lambda_{q\bf4}^2}{2\ m_{\bf4}^2}+2c_L^\dagger c_R^\dagger\frac{\lambda_{d\bf1}^2m_{\bf4}}{m_{\bf1}^3}+c_L c_R^\dagger\frac{\lambda_{q\bf4}^2m_{\bf1}}{m_{\bf4}^3}\right)
\right. \nonumber\\
&\ \ \ \ \ \ \ \ \ \ \ \ \ \ \ \left.-(y_{\bf4}-y_{\bf1})^2\lambda_{q\bf4}\lambda_{d\bf1}\frac{f}{\sqrt{2}}\left(\frac{c_R^\dagger}{m_{\bf1}m_{\bf4}^2}+\frac{c_L^\dagger}{m_{\bf1}^2m_{\bf4}}\right)
\right]
 \ , \nonumber
\end{align}
\begin{align}
& C_{d}^{h}\simeq L \frac{v}{\sqrt{2}}\left[-
\frac{\lambda_{q\bf4}\lambda_{d\bf4}}{f\ m_{\bf4}}+\frac{\lambda_{q\bf1}\lambda_{d\bf1}}{f\ m_{\bf1}}
+(y_{\bf4}-y_{\bf1})\left(\frac{2\lambda_{d\bf4}^2-\lambda_{q\bf4}^2}{2\ m_{\bf4}^2}+\frac{\lambda_{q\bf1}^2-2\lambda_{d\bf1}^2}{2\ m_{\bf1}^2}\right)
\right. \nonumber\\
&\ \ \ \ \ \ \ \ \ \ \ \ \ \ \ \left.+(y_{\bf4}-y_{\bf1})^2f\left(-\frac{\lambda_{q\bf1}\lambda_{d\bf1}}{m_{\bf1}^3}+\frac{\lambda_{q\bf4}\lambda_{d\bf4}}{m_{\bf4}^3}\right)\right]
 \ , \nonumber
\end{align}
\begin{align}
& C_{d}^{h\ddh}\simeq L \frac{v}{3\sqrt{2}}i\left[-
\lambda_{q\bf4}\lambda_{d\bf1}\left(\frac{c_L}{f\ m_{\bf4}}+\frac{c_R^\dagger}{f\ m_{\bf1}}\right)+\lambda_{q\bf1}\lambda_{d\bf4}\left(\frac{c_R}{f\ m_{\bf4}}+\frac{c_L^\dagger}{f\ m_{\bf1}}\right)
\right. \nonumber\\
&\ \ \ \ \ \ \ \ \ \ \ \ \ \ \ \ \ \ \ \left.+i(y_{\bf4}-y_{\bf1})^2\lambda_{q\bf4}\lambda_{d\bf1}\frac{f}{3\sqrt{2}}\left(\frac{c_R-c_R^\dagger}{m_{\bf1}m_{\bf4}^2}+\frac{c_L-c_L^\dagger}{m_{\bf1}^2m_{\bf4}}\right)\right] \ , \nonumber
\\
& C_{d}^{(\ddh)^2}\simeq L \frac{v}{3\sqrt{2}}\left[
2c_Lc_Rf\frac{\lambda_{q\bf4}\lambda_{d\bf4}}{m_{\bf1} m_{\bf4}^2}-2c_L^\dagger c_R^\dagger f\frac{\lambda_{q\bf1}\lambda_{d\bf1}}{m_{\bf4} m_{\bf1}^2}
+(y_{\bf4}-y_{\bf1})f^2\left(
c_Lc_R\frac{\lambda_{q\bf4}^2}{m_{\bf1} m_{\bf4}^3}+2c_L^\dagger c_R^\dagger\frac{\lambda_{d\bf1}^2}{m_{\bf4} m_{\bf1}^3}
\right)\right] \ .\label{eq-aproxloop-4}
\end{align}
where $L=g_{\rm SM}(16\pi^2 m^{(1)})^{-1}$. These contributions are in agreement with the estimate of Eq.~(\ref{Cd}).

The corrections to the $Z$ coupling can be computed by projecting $G^Z_{L,R}$ with the eigenvectors of the lightest mode and subtracting the SM coupling. Expanding in powers of $\lambda$ and $y$ we obtain:
\begin{eqnarray}
&(\hat G_R^Z)_{11}\simeq &-
\frac{g}{c_W}v^2\left[(y_{\bf4}-y_{\bf1})f\left(\frac{\lambda_{d{\bf1}}\lambda_{q{\bf4}}(c_R-c_R^\dagger)}{2\sqrt{2}m_{\bf4}^2m_{\bf1}}+\frac{\lambda_{d{\bf4}}\lambda_{q{\bf4}}}{2m_{\bf4}^3}\right)
+
(y_{\bf4}-y_{\bf1})^2f^2\frac{\lambda_{q{\bf4}}^2}{4m_{\bf4}^4}\right]
\ , \label{g1Rmchm}
\\
&(\hat G_L^Z)_{11}\simeq &
\frac{g}{c_W}v^2\left[
\frac{\lambda_{q{\bf4}}^2}{4m_{\bf4}^2}+\frac{\lambda_{q{\bf1}}^2}{4m_{\bf1}^2}+\frac{\lambda_{q{\bf4}}\lambda_{q{\bf1}}}{2\sqrt{2}m_{\bf4} m_{\bf1}}(c_L-c_L^\dagger)
\right.\nonumber\\
& &+
\left.(y_{\bf4}-y_{\bf1})f\left(\frac{\lambda_{d{\bf1}}\lambda_{q{\bf4}}(c_L-c_L^\dagger)}{2\sqrt{2}m_{\bf1}^2m_{\bf4}}-\frac{\lambda_{d{\bf1}}\lambda_{q{\bf1}}}{2m_{\bf1}^3}\right)
+
(y_{\bf4}-y_{\bf1})^2f^2\frac{\lambda_{d{\bf1}}^2}{4m_{\bf1}^4}\right]
\ . \label{g1Lmchm} 
\end{eqnarray}
Notice that the contribution of order ${\cal O}((y_{\bf4}-y_{\bf1})^0)$ is absent in $(\hat G_R^Z)_{11}$, since this coupling is protected by $P_C$-symmetry~\cite{Agashe:2006at}. Taking into account this cancellation, Eqs.~(\ref{g1Rmchm}) and~(\ref{g1Lmchm}) are in agreement with the estimate of Eq.~(\ref{CZ}).

Rotating to the mass basis, the composite 4-fermion operators generate a contribution to $C_{4f}$. Expanding in powers of $y$ and $\lambda$, we obtain for the mixed-chirality operator:
\begin{align}\label{C4f-mchm}
&C_{4f}\simeq C\left\{
f^2\frac{\lambda_{q{\bf4}}^2\lambda_{d{\bf1}}^2}{m_{{\bf4}}^2m_{{\bf1}}^2}\right.
+(y_{\bf4}-y_{\bf1})f v^2\left(-\frac{\lambda_{q{\bf1}}\lambda_{d{\bf1}}^3}{m_{{\bf1}}^5}+\frac{\lambda_{q{\bf4}}^3\lambda_{d{\bf4}}}{m_{{\bf4}}^5}-\frac{\lambda_{q{\bf1}}\lambda_{d{\bf1}}\lambda_{q{\bf4}}^2}{m_{{\bf1}}^3m_{{\bf4}}^2}+\frac{\lambda_{q{\bf4}}\lambda_{d{\bf4}}\lambda_{d{\bf1}}^2}{m_{{\bf1}}^2m_{{\bf4}}^3}\right)+\nonumber\\
&\ \ \ \ \ \ \ \ \ \ \ \ \ \left.+
(y_{\bf4}-y_{\bf1})^2f^2v^2\left[\frac{\lambda_{d{\bf1}}^4}{2m_{{\bf1}}^6}+\frac{\lambda_{q{\bf4}}^4}{2m_{{\bf4}}^6}+\frac{\lambda_{q{\bf4}}^2\lambda_{d{\bf1}}^2}{m_{{\bf4}}^2m_{{\bf1}}^2}\left(\frac{1}{m_{{\bf1}}^2}+\frac{1}{m_{{\bf4}}^2}\right)\right]\right\}
\ ,
\end{align}
where, for simplicity, we have taken: $C_{\br\bs}=\tilde C_{\br\bs}=C$. Eq.~(\ref{C4f-mchm}) leads to the estimate of Eq.~(\ref{C4f}).



\begin{thebibliography}{99}
\bibitem{Kaplan:1991dc}
  D.~B.~Kaplan,
  Nucl.\ Phys.\ B {\bf 365} (1991) 259.
  doi:10.1016/S0550-3213(05)80021-5

\bibitem{Grossman:1999ra}
  Y.~Grossman and M.~Neubert,
  Phys.\ Lett.\ B {\bf 474} (2000) 361
  doi:10.1016/S0370-2693(00)00054-X
  [hep-ph/9912408].

\bibitem{Huber:2000ie}
  S.~J.~Huber and Q.~Shafi,
  Phys.\ Lett.\ B {\bf 498} (2001) 256
  doi:10.1016/S0370-2693(00)01399-X
  [hep-ph/0010195].

\bibitem{Gherghetta:2000qt}
  T.~Gherghetta and A.~Pomarol,
  Nucl.\ Phys.\ B {\bf 586} (2000) 141
  doi:10.1016/S0550-3213(00)00392-8
  [hep-ph/0003129].

\bibitem{Contino:2004vy}
  R.~Contino and A.~Pomarol,
  JHEP {\bf 0411} (2004) 058
  doi:10.1088/1126-6708/2004/11/058
  [hep-th/0406257].

\bibitem{Agashe:2004cp}
  K.~Agashe, G.~Perez and A.~Soni,
  Phys.\ Rev.\ D {\bf 71} (2005) 016002
  doi:10.1103/PhysRevD.71.016002
  [hep-ph/0408134].

\bibitem{Agashe:2006iy}
  K.~Agashe, A.~E.~Blechman and F.~Petriello,
  Phys.\ Rev.\ D {\bf 74} (2006) 053011
  doi:10.1103/PhysRevD.74.053011
  [hep-ph/0606021].

\bibitem{Csaki:2008zd}
  C.~Csaki, A.~Falkowski and A.~Weiler,
  JHEP {\bf 0809} (2008) 008
  doi:10.1088/1126-6708/2008/09/008
  [arXiv:0804.1954 [hep-ph]].

\bibitem{Bauer:2009cf}
  M.~Bauer, S.~Casagrande, U.~Haisch and M.~Neubert,
  JHEP {\bf 1009} (2010) 017
  doi:10.1007/JHEP09(2010)017
  [arXiv:0912.1625 [hep-ph]].

\bibitem{Fitzpatrick:2007sa}
  A.~L.~Fitzpatrick, G.~Perez and L.~Randall,
  Phys.\ Rev.\ Lett.\  {\bf 100} (2008) 171604
  doi:10.1103/PhysRevLett.100.171604
  [arXiv:0710.1869 [hep-ph]].

\bibitem{Cacciapaglia:2007fw}
  G.~Cacciapaglia, C.~Csaki, J.~Galloway, G.~Marandella, J.~Terning and A.~Weiler,
  JHEP {\bf 0804} (2008) 006
  doi:10.1088/1126-6708/2008/04/006
  [arXiv:0709.1714 [hep-ph]].

\bibitem{Csaki:2008eh}
  C.~Csaki, A.~Falkowski and A.~Weiler,
  Phys.\ Rev.\ D {\bf 80} (2009) 016001
  doi:10.1103/PhysRevD.80.016001
  [arXiv:0806.3757 [hep-ph]].

\bibitem{Santiago:2008vq}
  J.~Santiago,
  JHEP {\bf 0812} (2008) 046
  doi:10.1088/1126-6708/2008/12/046
  [arXiv:0806.1230 [hep-ph]].

\bibitem{Redi:2011zi}
  M.~Redi and A.~Weiler,
  JHEP {\bf 1111} (2011) 108
  doi:10.1007/JHEP11(2011)108
  [arXiv:1106.6357 [hep-ph]].

\bibitem{Barbieri:2012uh}
  R.~Barbieri, D.~Buttazzo, F.~Sala and D.~M.~Straub,
  JHEP {\bf 1207} (2012) 181
  doi:10.1007/JHEP07(2012)181
  [arXiv:1203.4218 [hep-ph]].

\bibitem{Domenech:2012ai}
  O.~Domenech, A.~Pomarol and J.~Serra,
  Phys.\ Rev.\ D {\bf 85} (2012) 074030
  doi:10.1103/PhysRevD.85.074030
  [arXiv:1201.6510 [hep-ph]].

\bibitem{Bauer:2011ah}
  M.~Bauer, R.~Malm and M.~Neubert,
  Phys.\ Rev.\ Lett.\  {\bf 108} (2012) 081603
  doi:10.1103/PhysRevLett.108.081603
  [arXiv:1110.0471 [hep-ph]].

\bibitem{DaRold:2012sz}
  L.~Da Rold, C.~Delaunay, C.~Grojean and G.~Perez,
  JHEP {\bf 1302} (2013) 149
  doi:10.1007/JHEP02(2013)149
  [arXiv:1208.1499 [hep-ph]].

\bibitem{DaRold:2017dbr}
  L.~Da Rold and I.~A.~Davidovich,
  arXiv:1704.08704 [hep-ph].

\bibitem{Rattazzi:2008pe}
  R.~Rattazzi, V.~S.~Rychkov, E.~Tonni and A.~Vichi,
  JHEP {\bf 0812} (2008) 031
  doi:10.1088/1126-6708/2008/12/031
  [arXiv:0807.0004 [hep-th]].

\bibitem{Agashe:2008fe}
  K.~Agashe, T.~Okui and R.~Sundrum,
  Phys.\ Rev.\ Lett.\  {\bf 102} (2009) 101801
  doi:10.1103/PhysRevLett.102.101801
  [arXiv:0810.1277 [hep-ph]].

\bibitem{Matsedonskyi:2014iha}
  O.~Matsedonskyi,
  JHEP {\bf 1502} (2015) 154
  doi:10.1007/JHEP02(2015)154
  [arXiv:1411.4638 [hep-ph]].

\bibitem{Cacciapaglia:2015dsa}
  G.~Cacciapaglia, H.~Cai, T.~Flacke, S.~J.~Lee, A.~Parolini and H.~Serôdio,
  JHEP {\bf 1506} (2015) 085
  doi:10.1007/JHEP06(2015)085
  [arXiv:1501.03818 [hep-ph]].

\bibitem{Panico:2016ull}
  G.~Panico and A.~Pomarol,
  JHEP {\bf 1607} (2016) 097
  doi:10.1007/JHEP07(2016)097
  [arXiv:1603.06609 [hep-ph]].

\bibitem{Contino:2003ve}
  R.~Contino, Y.~Nomura and A.~Pomarol,
  Nucl.\ Phys.\ B {\bf 671} (2003) 148
  doi:10.1016/j.nuclphysb.2003.08.027
  [hep-ph/0306259].

\bibitem{Agashe:2004rs}
  K.~Agashe, R.~Contino and A.~Pomarol,
  Nucl.\ Phys.\ B {\bf 719} (2005) 165
  doi:10.1016/j.nuclphysb.2005.04.035
  [hep-ph/0412089].

\bibitem{Grojean:2013qca}
  C.~Grojean, O.~Matsedonskyi and G.~Panico,
  JHEP {\bf 1310} (2013) 160
  doi:10.1007/JHEP10(2013)160
  [arXiv:1306.4655 [hep-ph]].

\bibitem{Panico:2015jxa}
  G.~Panico and A.~Wulzer,
  Lect.\ Notes Phys.\  {\bf 913} (2016) pp.1
  doi:10.1007/978-3-319-22617-0
  [arXiv:1506.01961 [hep-ph]].

\bibitem{Redi:2013pga}
  M.~Redi,
  JHEP {\bf 1309} (2013) 060
  doi:10.1007/JHEP09(2013)060
  [arXiv:1306.1525 [hep-ph]].

\bibitem{KerenZur:2012fr}
  B.~Keren-Zur, P.~Lodone, M.~Nardecchia, D.~Pappadopulo, R.~Rattazzi and L.~Vecchi,
  Nucl.\ Phys.\ B {\bf 867} (2013) 394
  doi:10.1016/j.nuclphysb.2012.10.012
  [arXiv:1205.5803 [hep-ph]].

\bibitem{Agashe:2009di}
  K.~Agashe and R.~Contino,
  Phys.\ Rev.\ D {\bf 80} (2009) 075016
  [arXiv:0906.1542 [hep-ph]].

\bibitem{Mrazek:2011iu}
  J.~Mrazek, A.~Pomarol, R.~Rattazzi, M.~Redi, J.~Serra and A.~Wulzer,
  Nucl.\ Phys.\ B {\bf 853} (2011) 1
  [arXiv:1105.5403 [hep-ph]].

\bibitem{Olive:2016xmw}
  C.~Patrignani {\it et al.} [Particle Data Group],
  Chin.\ Phys.\ C {\bf 40} (2016) no.10,  100001.
  doi:10.1088/1674-1137/40/10/100001

\bibitem{Gedalia:2009ws}
  O.~Gedalia, G.~Isidori and G.~Perez,
  Phys.\ Lett.\ B {\bf 682} (2009) 200
  doi:10.1016/j.physletb.2009.10.097
  [arXiv:0905.3264 [hep-ph]].

\bibitem{Buras:2011ph}
  A.~J.~Buras, C.~Grojean, S.~Pokorski and R.~Ziegler,
  JHEP {\bf 1108} (2011) 028
  doi:10.1007/JHEP08(2011)028
  [arXiv:1105.3725 [hep-ph]].

\bibitem{Agashe:2006at}
  K.~Agashe, R.~Contino, L.~Da Rold and A.~Pomarol,
  Phys.\ Lett.\ B {\bf 641} (2006) 62
  doi:10.1016/j.physletb.2006.08.005
  [hep-ph/0605341].

\bibitem{Contino:2006nn}
  R.~Contino, T.~Kramer, M.~Son and R.~Sundrum,
  JHEP {\bf 0705} (2007) 074
  doi:10.1088/1126-6708/2007/05/074
  [hep-ph/0612180].

\bibitem{Agashe:2008uz}
  K.~Agashe, A.~Azatov and L.~Zhu,
  Phys.\ Rev.\ D {\bf 79} (2009) 056006
  [arXiv:0810.1016 [hep-ph]].

\bibitem{Giudice:2007fh}
  G.~F.~Giudice, C.~Grojean, A.~Pomarol and R.~Rattazzi,
  JHEP {\bf 0706} (2007) 045
  doi:10.1088/1126-6708/2007/06/045
  [hep-ph/0703164].

\bibitem{DeSimone:2012fs}
  A.~De Simone, O.~Matsedonskyi, R.~Rattazzi and A.~Wulzer,
  JHEP {\bf 1304} (2013) 004
  doi:10.1007/JHEP04(2013)004
  [arXiv:1211.5663 [hep-ph]].

\bibitem{Contino:2011np}
  R.~Contino, D.~Marzocca, D.~Pappadopulo and R.~Rattazzi,
  JHEP {\bf 1110} (2011) 081
  doi:10.1007/JHEP10(2011)081
  [arXiv:1109.1570 [hep-ph]].

\bibitem{Panico:2011pw}
  G.~Panico and A.~Wulzer,
  JHEP {\bf 1109} (2011) 135
  doi:10.1007/JHEP09(2011)135
  [arXiv:1106.2719 [hep-ph]].

\bibitem{Azatov:2013ura}
  A.~Azatov, R.~Contino, A.~Di Iura and J.~Galloway,
  Phys.\ Rev.\ D {\bf 88} (2013) no.7,  075019
  doi:10.1103/PhysRevD.88.075019
  [arXiv:1308.2676 [hep-ph]].

\bibitem{Antipin:2014mda}
  O.~Antipin, S.~De Curtis, M.~Redi and C.~Sacco,
  Phys.\ Rev.\ D {\bf 90} (2014) no.6,  065016
  doi:10.1103/PhysRevD.90.065016
  [arXiv:1407.2471 [hep-ph]].

\end{thebibliography}
\end{document}